\newcommand{\diracslash}[1]{#1\llap{/\kern2pt}}

\newcommand{\be}{\begin{equation}}
\newcommand{\ee}{\end{equation}}
\newcommand{\bea}{\begin{eqnarray}}
\newcommand{\eea}{\end{eqnarray}}
\newcommand{\ba}[1]{\begin{array}{#1}}
\newcommand{\ea}{\end{array}}

\newcommand{\bt}{\begin{tabular}}
\newcommand{\et}{\end{tabular}}

\newcommand{\beas}{\begin{eqnarray*}}
\newcommand{\eeas}{\end{eqnarray*}}

\RequirePackage{lineno}

\documentclass[preprint,prd,aps,floats,nofootinbib,floatfix]
 {revtex4-2}

\DeclareSymbolFont{rsfs}{U}{rsfs}{m}{n}
\DeclareSymbolFontAlphabet{\mathrsfs}{rsfs}

\usepackage{amsmath}
\usepackage{graphicx}
\usepackage[mathscr]{eucal}
\usepackage{multirow}
\usepackage{graphicx,epstopdf}
\usepackage{graphicx}
\usepackage{subcaption}
\usepackage{cancel}

\usepackage{setspace}
\usepackage[utf8]{inputenc}
\usepackage{csquotes}
\usepackage[english]{babel}
\usepackage{hyperref}
\usepackage{upgreek}
\hypersetup{
    colorlinks=true,
    linkcolor=blue,
    filecolor=magenta,     
    citecolor=blue, 
    urlcolor=blue,
}
\usepackage{cleveref}
\usepackage{caption}
\usepackage{relsize}
\begin{document}
\setstretch{1.5}
%\begin{linenumbers}
%\begin{linenumbers}
% \title{$J/\psi$ mass shift in asymmetric nuclear medium and $J/\psi$-nuclei: Impact of $D D$ and $DD^*$  meson loops}
\title{A study of $J/\psi$ mass shift and bound states: Impact of $D D$ and $DD^*$  meson loops}
%%%%%%%%%%%%%%%%%%%%%%%%%%%%%%%%%%%%%%%%%%%%%%%%%%%%%%%%%%%%%%%%%
\author{Manpreet Kaur}
\email{ranapreeti803@gmail.com}
\author{Arvind Kumar}
\email{kumara@nitj.ac.in}
\affiliation{Department of Physics, Dr. B R Ambedkar National Institute of Technology, 
	Jalandhar -- 144008, Punjab, India}

%
%
%\affiliation{Department of Physics, Dr. B R Ambedkar National Institute of Technology Jalandhar, 144008, Punjab, India}
\def\be{\begin{equation}}
\def\ee{\end{equation}}
\def\bearr{\begin{eqnarray}}
\def\eearr{\end{eqnarray}}
\def\zbf#1{{\bf {#1}}}
\def\bfm#1{\mbox{\boldmath $#1$}}
\def\hf{\frac{1}{2}}
\def\kp{\zbf k+\frac{\zbf q}{2}}
\def\km{-\zbf k+\frac{\zbf q}{2}}
\def\hwo{\hat\omega_1}
\def\hwt{\hat\omega_2}
\begin{abstract}
% We explore the in-medium modification of the $J/\psi$ meson mass in asymmetric nuclear matter at zero and finite temperatures using an effective Lagrangian via the contribution of $DD$ and $DD^*$ meson loops. The medium dependence of $D$ meson masses is calculated by the hadronic chiral SU(3) model in which scalar condensates are evaluated and are employed as input in the QCD sum rule approach. We found that with increase in baronic density we get negative mass shift and This negative mass shift implies that $J/\psi$ is attracted to nuclear mean fields.
% The $J/\psi$ self-energy is evaluated via the contribution of $DD$ and $DD^*$ meson loops, with medium effects incorporated through $D$ and $D^*$ meson masses computed in the hadronic chiral SU(3) model and QCD sum rules.
We investigate the modification of the $J/\psi$ meson mass in asymmetric nuclear matter at zero and finite temperatures employing an effective Lagrangian approach that considers the contributions of $DD$ and $DD^*$ meson loops. The medium dependence of $D$ meson masses is determined using the hadronic chiral SU(3) model, where scalar condensates are calculated and subsequently utilized in the QCD sum rules approach. Our findings indicate that an increase in baryonic density results in a negative mass shift of the $J/\psi$ meson. This suggests that the $J/\psi$ meson is attracted to nuclear mean fields indicating the possibility of the formation of meson-nucleus bound states. Moreover, we have also determined the binding energy and absorption decay width of the $J/\psi$ meson for both the ground and excited states of $\text{O}^{16}$, $\text{Ca}^{40}$, $\text{Zr}^{90}$, and $\text{Pb}^{208}$ nuclei. These results are expected to contribute to the understanding of experimental data from upcoming studies at Jefferson lab and the facility for antiproton and ion research, where low-momentum charmed mesons can be produced and examined within nuclei.

\end{abstract}
\maketitle
 \newpage
\section{Introduction\label{secintro}}

A quantitative description of strongly interacting matter is still an open challenge, especially for hadrons immersed in hot and dense matter, where they serve as probes of chiral symmetry. The study of heavy-light and heavy-heavy meson interactions with nucleons is essential for exploring the properties of hadronic matter, both in a vacuum and under conditions of extreme temperature and density \cite{Hosaka:2016ypm, Krein:2016fqh, Leupold:2009kz,Hayano:2008vn, Cobos-Martinez:2025iqg}.
The suppression of $J/\psi$ mesons was proposed by Matsui and Satz as evidence for quark-gluon plasma (QGP) formation \cite{Matsui:1986dk}. Early measurements by the NA38 and NA50 collaborations investigated this effect in heavy ion collisions \cite{NA38:1989yxm, NA38:1994yau, NA50:2001xnx}. In a deconfined medium, color Debye screening reduces the binding strength between $c \bar c$ pairs in extreme conditions, causing the $J/\psi$ mesons to dissociate. The liberated charm quarks can subsequently produce D mesons through hadronization \cite{Bazavov:2020teh, Calucci:2006bs, Shao:2020kgj}.
Open heavy-flavor mesons, such as $D$ mesons, exhibit a greater sensitivity to the surrounding medium than heavy quarkonia like $J/\psi$ meson. This increased sensitivity is due to their interaction with light-quark condensates, which undergo significant changes with variations in temperature and density. As a result, these mesons are valuable tools for probing the deconfined quantum chromodynamics (QCD) medium, which is crucial for understanding the conditions of the early universe and the nature of compact stars.
At the facility for antiproton and ion research (FAIR), the compressed baryonic matter (CBM) collaboration seeks to explore the characteristics of hadrons within dense nuclear matter \cite{senger2012compressed, Agarwal:2023otg, Durante:2019hzd}. Concurrently, the PANDA (anti-Proton Annihilation at Darmstadt) collaboration is set to offer valuable insights into $D$ meson and charmonium spectroscopy, alongside advancing the understanding of strangeness physics and electromagnetic processes \cite{Tomasi-Gustafsson:2018fwm, Belias:2020zwx, Durante:2019hzd}. Moreover, the PANDA experiment is designed to probe bound states extending to the charm sector, with the potential to explore heavier quark states, subject to the constraints of experimental capabilities \cite{Frankfurt:2019uvk, Durante:2019hzd}. The experiments, such as the continuous electron beam accelerator Facility (CEBAF) at Jefferson Lab delve into low-momentum meson production within nuclei, offering insight into the existence and behavior of charm quarks \cite{Adderley:2024czm}. The J-PARC-E29 collaboration intends to investigate the modification of meson masses by examining meson-nucleus bound states, which are produced through the process of antiproton annihilation \cite{Tanida:2016ryv}. The experimental determination of in-medium mass modifications remains a difficult challenge. Consequently, more comprehensive and detailed theoretical studies are essential to enhance our understanding of these effects.

Significant progress has been made in the theoretical study of heavy-flavor mesons and heavy-quarkonium interactions in hadronic media over the past several decades using numerous nonperturbative theoretical frameworks \cite{Eichten:1978tg,Karsch:1987pv, Mocsy:2005qw, Bonati:2015dka, Hayashigaki:2000es, Kim:2000kj, Klingl:1998sr, Kumar:2010hs, Morita:2010pd, Hilger:2011cq, Zschocke:2011aa, Chhabra:2016vhp, Kumar:2015fca, Kumar:2020igz, Tolos:2004yg, Tolos:2007vh, Hofmann:2005sw, Molina:2009zeg, Neubert:1993mb, Yasui:2012rw, Yasui:2013iga, Tsushima:1998ru, Tsushima:2002cc, Saito:1994ki, Krein:2010vp, Krein:2017usp, Tsushima:2011fg, Zeminiani:2020aho, Lee:2000csl, Krein:2010vp, Krein:2017usp, Tsushima:2011fg, Zeminiani:2020aho, Kaidalov:1992hd, Luke:1992tm, TarrusCastella:2018php, Sibirtsev:2005ex}. Early research relied on phenomenological potential models \cite{Eichten:1978tg,Karsch:1987pv, Mocsy:2005qw, Bonati:2015dka} to describe the binding and interaction dynamics between heavy mesons and nucleons. As the field progressed, more systematic approaches, such as QCD sum rules, were adopted to analyze the medium modifications of heavy-flavor meson properties by relating them to changes observed in quark and gluon condensates \cite{Hayashigaki:2000es, Kim:2000kj, Klingl:1998sr, Kumar:2010hs, Morita:2010pd, Hilger:2011cq, Zschocke:2011aa, Chhabra:2016vhp, Kumar:2015fca, Kumar:2020igz}. Coupled-channel approaches \cite{Tolos:2004yg, Tolos:2007vh, Hofmann:2005sw, Molina:2009zeg} and heavy meson effective theories \cite{Neubert:1993mb, Yasui:2012rw, Yasui:2013iga} provide a more dynamical description by incorporating multi-channel effects and heavy quark symmetry, respectively. Models based on quark–meson coupling (QMC) \cite{Tsushima:1998ru, Tsushima:2002cc, Saito:1994ki, Krein:2010vp, Krein:2017usp, Tsushima:2011fg, Zeminiani:2020aho} have also been utilized to examine the interactions of heavy mesons with nucleons in dense matter. Moreover, other theoretical approaches for quarkonium–nucleon interactions include charmed meson loop contributions \cite{Lee:2000csl, Krein:2010vp, Krein:2017usp, Tsushima:2011fg, Zeminiani:2020aho}, color polarizability effects \cite{Peskin:1979va, Kharzeev:1995ij, Sibirtsev:2005ex}, and long-range van der Waals–type interactions \cite{Lee:2000csl, Kaidalov:1992hd, Luke:1992tm, TarrusCastella:2018php, Sibirtsev:2005ex}. Lattice QCD simulations have recently offered significant insights into charmonium–nucleon interactions and the potential for forming bound systems with nuclei, although many investigations still rely on heavier pion masses \cite{Beane:2014sda, Alberti:2016dru}.

Previous theoretical studies have reported several estimates for the behavior of the $J/\psi$ meson in nuclear medium. The QCD sum rules studies indicate a relatively small mass reduction of $4$ to $7\,\mathrm{MeV}$ in the $J/\psi$ meson \cite{Klingl:1998sr, Hayashigaki:1998ey, Kim:2000kj}. On the other hand, analyses based on color polarizability predict a larger decrease in $J/\psi$ meson mass, about $21$ MeV \cite{Sibirtsev:2005ex}.  Additionally, calculations carried out within the QMC model show that medium modifications of the $DD$, $DD^{*}$, and $D^{*}D^{*}$ meson loops at normal nuclear matter density lead to a $J/\psi$ mass shift ranging from $-16$ to $-24\,\mathrm{MeV}$ \cite{Krein:2010vp}. The unexpected impact of the heavier $D^{*}D^{*}$ meson loop on the $J/\psi$ self-energy was observed, the authors have revised their prediction of the $J/\psi$ mass shift by incorporating the contribution from the DD meson loop only, as detailed in Ref. \cite{Krein:2017usp, CobosMartinez:2021bia}. The unexpected finding may be linked to two factors: the assumption of identical coupling strengths for the $DD$, $DD^{*}$, and $D^{*}D^{*}$ meson loops and uncertainties in the formulation of the effective Lagrangian \cite{Zeminiani:2020aho, Krein:2017usp}. The chiral effective model \cite{Papazoglou:1998vr, bardeen1969some,weinberg1968nonlinear, Zschiesche:2003qq} has been widely used to explore the behavior of heavy quark systems in a dense hadronic environment, including the studies of the in-medium masses of charmonium and bottomonium states \cite{Kumar:2010gb, Mishra:2014gea} and open heavy-flavor mesons \cite{Kumar:2010gb, Mishra:2014gea, Mishra:2003se, Pathak:2014nfa}. In order to examine the decay mechanisms of quarkonium into open heavy-flavor mesons within hadronic matter, a combined approach of the chiral SU(3) model and the $^3 P_0$ pair-creation model was utilized \cite{Barnes:1996ff, LeYaouanc:1972vsx}. Investigating $J/\psi$ photoproduction in the threshold kinematic region is crucial for providing input to theoretical models that calculate the gluon structure of the proton. This includes the generalized parton distribution (GPD), the mass radii of proton, and the contribution of the trace anomaly to the proton mass \cite{Lee:2022ymp, Kim:2024lxc, He:2024lry}.
% In Ref. \cite{Chhabra:2020dsr}, the authors examined the mass shift of $J/\psi$ within a symmetric nuclear medium and a strange medium, focusing solely on the contributions from the $D$ meson loop.  In this study, we extending this work for study investigate the in-medium modification of the $J/\psi$ meson mass in asymmetric nuclear matter at both zero and finite temperatures, employing an effective Lagrangian approach that includes contributions from $DD$ and $DD^*$ meson loops. 
In Ref. \cite{Chhabra:2020dsr}, the in-medium mass shift of the $J/\psi$ meson was analyzed in symmetric nuclear and strange matter, taking into account only the pseudoscalar $DD$ meson loop contribution.

The study of attractive meson-nucleus interaction would provide evidence for the negative mass shift of mesons and the role of gluon forces in nuclei \cite{Hosaka:2016ypm, Krein:2016fqh, Leupold:2009kz,Hayano:2008vn, piAF:2022gvw, Metag:2017yuh,Cobos-Martinez:2025iqg}. Facilities such as JLab and FAIR will be able to produce low-momentum $J/\psi$-nucleus bound states \cite{Frankfurt:2019uvk, Durante:2019hzd, Adderley:2024czm}.
% Within the relativistic mean field model, Meson-nucleon interactions can be better understood by examining mesic-nuclei bound states. 
% In literature, many methods are used to determine the binding energies of mesic-nuclei bound states. 
The second-order Stark effect \cite{Szmytkowski:2018iog}, which results from chromo-electric polarizability, plays a crucial role in determining the mass shift of charmonium in nuclear matter. Assuming a charm quark of infinite mass, predictions indicated a binding energy of approximately 10~MeV for the $J/\psi$ in nuclear matter and 70~MeV for the excited charmonium states \cite{Peskin:1979va, Luke:1992tm}. According to Ref. \cite{Lee:2000csl}, the binding energy for $J/\psi$ in nuclear matter was found to be 8~MeV. In contrast, when the same theoretical method was applied with a finite charm quark mass and realistic charmonium bound state wave functions, the binding energy for the excited charmonium states exceeded 100~MeV. Coupled-channel approaches have been successful in identifying $D$ meson bound states, which hold significant importance for the PANDA and CBM experiments at FAIR \cite{Garcia-Recio:2010fiq, Yamagata-Sekihara:2015ebw}. Similar investigations have been conducted for various mesons such as $\rho$, $\omega$,$\eta$, $\phi$, and $J/\psi$ mesons \cite{Schadmand:2010bi, Tsushima:2011kh,Saito:1997ae, Cobos-Martinez:2020ynh,CobosMartinez:2021bia, Cobos-Martinez:2022fmt, Cobos-Martinez:2025iqg}. In Refs. \cite{Tsushima:2011kh, Mondal:2024vyt}, the energies of heavy meson-nucleus bound states were determined by solving the Klein-Gordon equations, using real potentials that were calculated self-consistently within the quark meson field model. Most recently, Ref. \cite{CobosMartinez:2021bia} presented updated results on $J/\psi$-nucleus and $\eta_c$-nucleus bound states, incorporating an intermediate $D D^*$ state in the calculation of the $\eta_c$ self-energy.

In this study, we explore the modification of the $J/\psi$ mass in isospin asymmetric nuclear matter at both zero and finite temperatures, employing an effective Lagrangian approach that includes $DD$ and $DD^*$ meson loops. Additionally, we examine the binding energy and absorption decay width of the $J/\psi$ meson for the ground and excited states of the $\text{O}^{16}$, $\text{Ca}^{40}$, $\text{Zr}^{90}$, and $\text{Pb}^{208}$ nuclei by solving Klein-Goldon equation. The medium dependence of $D$ meson masses is determined using the hadronic chiral SU(3) model, where scalar condensates are calculated and subsequently utilized in the QCD sum rules. Within the chiral SU(3) model, the dilaton field, $\chi$, is introduced primarily to replicate the gluon dynamics of QCD and to account for the effects of QCD scale symmetry breaking at the hadronic level \cite{Papazoglou:1998vr, Kaur:2024cfm}. 
The finite isospin asymmetry of the medium is explained by the vector-isovector field $\rho$ and the scalar isovector field $\delta$.
% This characteristic makes it especially valuable for examining charmonium and heavy mesons, whose masses are strongly influenced by the gluon condensate. These condensates are used directly as inputs to QCD sum rules.      

This manuscript is structured as follows. In Sec. \ref{chiral model}, we introduce the chiral SU(3) hadronic model and present the relevant results. Sec. \ref{QCDSR} describes the QCD sum rules approach and the obtained results for the $D$ meson masses. Sec. \ref{ELA} focuses on the effective Lagrangian framework and the corresponding results of medium modifications of the $J/\psi$ meson. We present an overview of mesic-nucleus potentials, along with a discussion on the binding energy and absorption decay width of mesons in nuclei, in Sec. \ref{sec_bound_state}.
The main findings are summarized in Section \ref{summary}.

\section{The chiral SU(3) hadronic model}
\label{chiral model}
The chiral SU(3) hadronic mean-field model provides an effective theoretical framework for analyzing the non-perturbative dynamics of strong interactions, founded on a nonlinear realization of chiral symmetry and the effects of broken scale invariance \cite{Papazoglou:1998vr,bardeen1969some,weinberg1968nonlinear}. The effective Lagrangian density associated with this model is represented as \cite{Papazoglou:1998vr, Mishra:2014rha, Kaur:2024cfm}

\begin{equation}
\mathcal{L}_{\text{chiral}} = \mathcal{L}_{\text{kin}} + \sum_{M=P,S,V,A} \mathcal{L}_{\text{BM}} + \mathcal{L}_{\text{Vec}} + \mathcal{L}_{0} + \mathcal{L}_{\text{ESB}},
\end{equation}
where \( \mathcal{L}_{\text{kin}} \) denotes the terms related to kinetic energy,  
\( \mathcal{L}_{\text{BM}} \) characterizes the interactions between baryons and mesons,  
\( \mathcal{L}_{\text{Vec}} \) represents the mass term for the vector mesons as well as their self-interactions,  
\( \mathcal{L}_{0} \) is the self-interaction term of scalar mesons, and the final term \( \mathcal{L}_{\text{ESB}} \) accounts for explicit symmetry breaking.
% where \( \mathcal{L}_{\text{kin}} \) represents kinetic energy terms,  
% \( \mathcal{L}_{\text{BM}} \) describes baryon-meson interactions,  
% \( \mathcal{L}_{\text{Vec}} \) determines the vector meson mass via its interactions with scalar mesons and includes quartic self-interaction terms,  
% \( \mathcal{L}_{0} \) contains meson-meson interaction terms that lead to spontaneous chiral symmetry breaking, and last term \( \mathcal{L}_{\text{ESB}} \) incorporates explicit symmetry breaking.
 The interaction between baryons and mesons can be described through the exchange of scalar fields $\sigma$, $\zeta$, $\delta$, as well as vector fields $\omega$, and $\rho$.
The mean-field approximation is utilized to examine the characteristics of hadronic matter by replacing the quantum field operators with their classical expectation values for scalar and vector fields, as these fields are anticipated to exhibit minimal fluctuations in dense environments \cite{Papazoglou:1998vr, Kaur:2024cfm}. Applying the Euler–Lagrange equation of motion, $\frac{\partial \mathcal{L}}{\partial \phi_i}
	-
	\partial_\mu \left(
	\frac{\partial \mathcal{L}}{\partial (\partial_\mu \phi_i)}
	\right)
	= 0,$ for given field $\phi_i$,
% \begin{equation}
% 	\frac{\partial \mathcal{L}}{\partial \phi_i}
% 	-
% 	\partial_\mu \left(
% 	\frac{\partial \mathcal{L}}{\partial (\partial_\mu \phi_i)}
% 	\right)
% 	= 0,
% \end{equation}
we obtain a set of coupled equations for the scalar fields \( \sigma \), \( \zeta \), \( \delta \), the dilaton field \( \chi \), and the vector fields \( \omega \) and \( \rho \)~\cite{Chhabra:2016vhp,Kaur:2024cfm, Kumar:2010gb}.
To determine the in-medium masses of \( D \) and $D^*$ mesons using QCD sum rules \cite{Wang:2015uya, Mishra:2014rha}, it is essential to evaluate the light quark condensates \( \langle \bar{u}u \rangle \) and \( \langle \bar{d}d \rangle \), along with the scalar gluon condensate \( \left\langle \frac{\alpha_s}{\pi} G^a_{\mu\nu} G^{\mu\nu}_a \right\rangle \).
In the chiral SU(3) model, the scalar light quark condensates are linked to explicit chiral symmetry breaking through the following relation \cite{Kumar:2019axp},
\begin{equation}
\sum_i m_i \left\langle \bar{q}_i q_i \right\rangle_{\rho_B} = -\mathcal{L}_{\mathrm{ESB}}.
\end{equation}
The scalar light quark condensates are given by
\begin{equation}
\label{eq_uu}
\langle \bar{u} u \rangle_{\rho_B} = \frac{1}{m_u} \left(\frac{\chi}{\chi_0}\right)^2 
\left[\frac{1}{2} m_\pi^2 f_\pi (\sigma + \delta)\right],
\end{equation}
and
\begin{equation}
\label{eq_dd}
\langle \bar{d} d \rangle_{\rho_B} = \frac{1}{m_d} \left(\frac{\chi}{\chi_0}\right)^2 
\left[\frac{1}{2} m_\pi^2 f_\pi (\sigma - \delta)\right],
\end{equation}
here, $m_u$ and $m_d$ denote the masses of the up and down quarks, i.e., $m_u=5$~MeV and $m_d=7$~MeV.
\begin{table}[ht]
\centering
\caption{The coupling constants and other parameters used in the current work.}
\label{tab:constants}
\renewcommand{\arraystretch}{1.3} % Adds vertical space for subscripts and fractions
\begin{tabular}{|l|l|l|l|l|l|}
\hline
Parameter & Value & Parameter & Value & Parameter & Value \\ \hline
$k_0$ & 2.53 & $\sigma_0$ (MeV) & -93.3 & $g_{\sigma N}$ & 10.6 \\ \hline
$k_1$ & 1.35 & $\zeta_0$ (MeV) & -106.76 & $g_{\zeta N}$ & -0.46 \\ \hline
$k_2$ & -4.77 & $\chi_0$ (MeV) & 409.76 & $g_{\delta N}$ & 2.34 \\ \hline
$k_3$ & -2.77 & $d$ & 0.064 & $g_{\omega N}$ & 13.42 \\ \hline
$k_4$ & -0.21 & $g_4$ & 79.9 & $g_{\rho N}$ & 5.48 \\ \hline
$m_{\pi}$ (MeV) & 139 & $m_K$ (MeV) & 498 & $f_{\pi}$ (MeV) & 93.3 \\ \hline
$f_K$ (MeV) & 122.14 & $\rho_0$ ($\text{fm}^{-3}$) & 0.15 & $m_N$ (MeV) & 939 \\ \hline
$m_{\omega}$ (MeV) & 783 & $m_{\rho}$ (MeV) & 770 & $m_{\phi}$ (MeV) & 1020 \\ \hline
\end{tabular}
\label{key_para}
\end{table}

In QCD, the breaking of scale invariance results in the energy-momentum tensor having a nonzero trace, which is directly linked to the scalar gluon condensate and given as \cite{Papazoglou:1998vr, Weinberg:1968de}
\begin{figure*}[h]
	\centering
	\includegraphics[width=16cm]{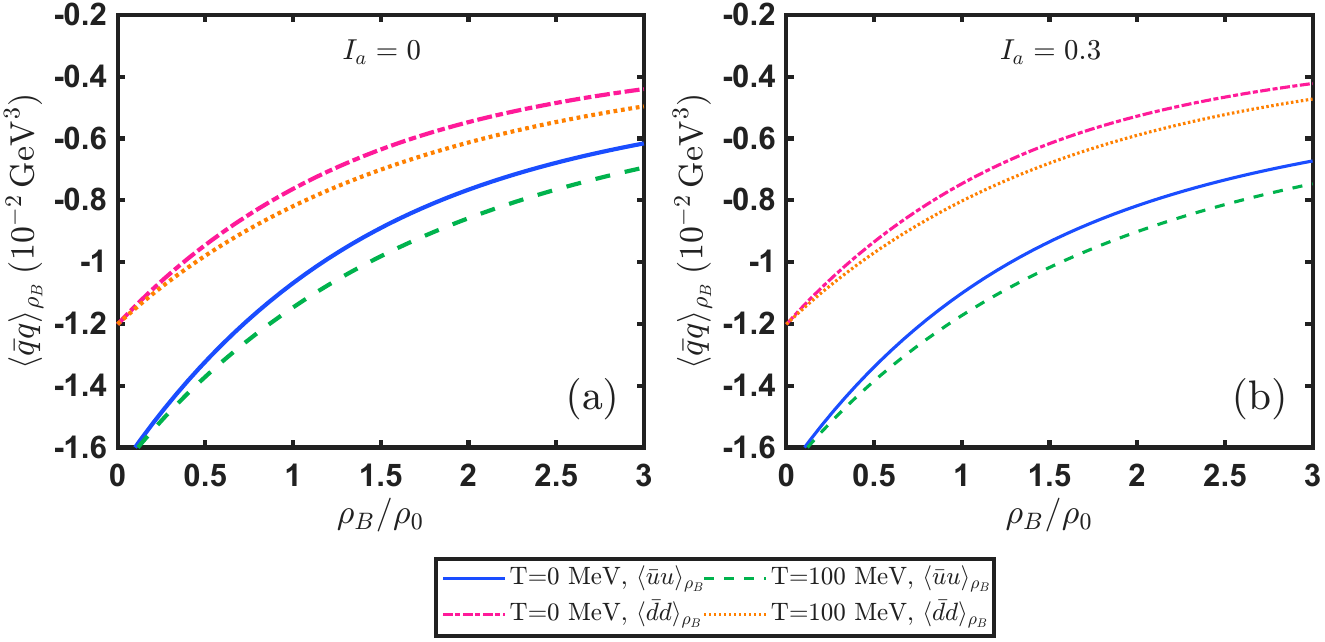}
	\caption{The up \( \langle u\bar{u} \rangle_{\rho_B} \) and down \( \langle d\bar{d} \rangle_{\rho_B} \) condensates are analyzed as function of the baryon density ratio \( \rho_B/\rho_0 \) in isospin asymmetric nuclear matter, considering isospin asymmetric parameter \( I_a = 0 \) and 0.3 and temperatures $T=0$ and 100 MeV.}
	\label{fig_quark_con}
\end{figure*} 
\begin{figure*}[h]
	\centering
	\includegraphics[width=16cm]{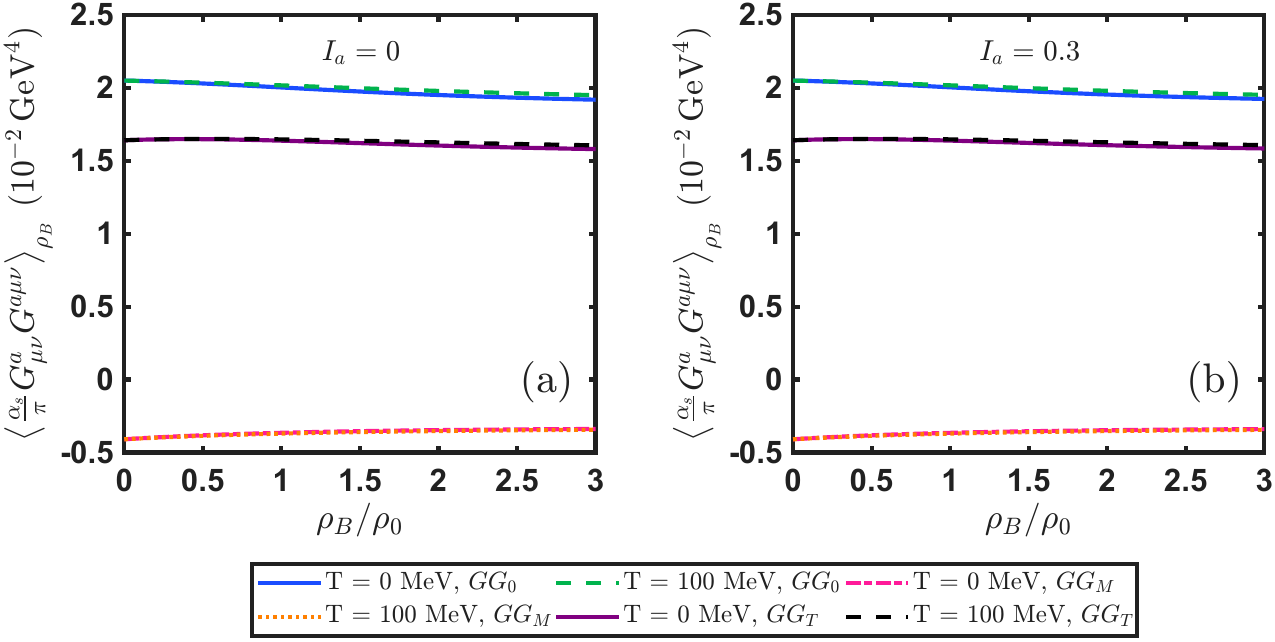}
	\caption{The individual and total terms of gluon condensates ($GG_0$, $GG_M$, and $GG_T$)  are examined as function of the baryon density ratio \( \rho_B/\rho_0 \) within nuclear matter, considering \( I_a = 0 \) and 0.3 and temperatures $T=0$ and 100 MeV.}
	\label{fig_gluon_con}
\end{figure*}
\small{
\begin{equation}
	\label{gluon_con}
\begin{aligned}
\left\langle \frac{\alpha_s}{\pi} G_{\mu \nu}^a G^{a \mu \nu} \right\rangle_{\rho_B}
&= \frac{8}{9} \Bigg[
(1 - d)\, \chi^4 + \left(\frac{\chi}{\chi_0}\right)^2 \\
&\quad
\Big(
m_\pi^2 f_\pi \sigma 
+ \big( \sqrt{2} m_K^2 f_K 
- \tfrac{1}{\sqrt{2}} m_\pi^2 f_\pi \big)\zeta
\Big)
\Bigg],
\end{aligned}
\end{equation}
}
where \( d = \frac{2}{11} \) serves as a parameter associated with the QCD beta function \cite{Papazoglou:1998vr}.
% In this present work, we discuss the results of the in-medium properties of isospin pseudoscalar \(D\) mesons (\(D^+\), \(D^0\)), vector \(D^*\) mesons (\(D^{*+}\), \(D^{*0}\)) and $J/\psi$ in an isospin-asymmetric nuclear medium. 
In the present work, we examine the chiral model's predictions for determining the in-medium values of the chiral condensates, particularly emphasis on the light quark condensates $\langle \bar{u} u \rangle_{\rho_B}$, $\langle \bar{d} d \rangle_{\rho_B}$, and the gluon condensate $\left\langle \frac{\alpha_s}{\pi} G_{\mu \nu}^a G^{a \mu \nu} \right\rangle_{\rho_B}$. We examine the impact of the isospin asymmetry of the medium, defined as $I_a = - \sum_i\frac{I_{3i}\rho_{i}}{2\rho_B}$, where $\rho_B$ represents the total baryonic density, $\rho_i$ is the nucleon number density, and $I_{3i}$ is the third component of isospin of nucleons. In Figs. \ref{fig_quark_con} and \ref{fig_gluon_con}, the results for these condensates are illustrated for isospin asymmetry values $I_a = 0$ and $0.3$, and at temperatures $T = 0$ and $100$~MeV.
% These condensates are modulated by the scalar fields \(\sigma\), \(\zeta\), \(\delta\), and \(\chi\), whose values are determined under varying medium conditions such as baryon density, temperature, and isospin asymmetry.
% The output from condensates are then used as inputs in QCD sum rules (QCDSR) to evaluate the effective masses of the  pseudoscalar \(D\) (\(D^+\), \(D^0\)), vector \(D^*\) (\(D^{*+}\), \(D^{*0}\)) mesons and $J/\psi$ in an isospin-asymmetric nuclear medium, which are discussed in the section ~\ref{sec:mass_modification}. All calculations are performed at nuclear saturation density (\(\rho_0 = 0.15\,\mathrm{fm}^{-3}\)). 
% All the important parameters (vacuum mass values and coupling constants), which we used in chiral model are listed in Table~\ref{tab:parameters}. 
The condensate outputs are subsequently utilized as input in QCD sum rules to determine the effective masses of the pseudoscalar \(D\) (\(D^+\), \(D^0\)) and vector \(D^*\) (\(D^{*+}\), \(D^{*0}\)) mesons within an isospin-asymmetric nuclear medium, as elaborated in Sec ~\ref{QCDSR}. 
%These computations are conducted at nuclear saturation density (\(\rho_0 = 0.15\,\mathrm{fm}^{-3}\)). 
The key parameters, including vacuum masses of mesons and coupling constants of mesonic fields with baryons within the nuclear medium used in this work, are listed in Table \ref{key_para}, and the fitting of parameters is detailed in Refs. \cite{Papazoglou:1998vr, Kaur:2024cfm}.
In Fig.~\ref{fig_quark_con}(a), for \(I_a = 0\), the results indicate that the magnitude of condensates 
\(\langle \bar{u} u \rangle_{\rho_B}\) and \(\langle \bar{d} d \rangle_{\rho_B}\) (of the order of \(10^{-2}\ \text{GeV}^{3}\)) decrease as the baryonic density increases at a given temperature. 

% The reduction in the magnitude of condensates is more significant for the $d$ quark than for the $u$ quark.
% which corresponds to the vacuum masses of these quarks. This pattern is explained by the partial restoration of chiral symmetry at high temperatures
As the temperature rises from zero to 100 MeV, there is a less pronounced drop in the magnitude of quark condensates than the value observed at $T=0$~MeV at a given $\rho_B /\rho_0$ as illustrated in Fig. \ref{fig_quark_con}(a). For instance, the value of \(\langle \bar{u} u \rangle_{\rho_B}\) (\(\langle \bar{d} d \rangle_{\rho_B}\)) is found to be -1.10 (-0.74) $\times 10^{-2}\,\text{GeV}^3 $ at a temperature $T=0$~MeV, However, when the temperature reaches $T=100$~MeV, these values change to -1.17 (-0.80)$\times 10^{-2}\,\text{GeV}^3$, at nuclear saturation density  (\(\rho_0= 0.15\,\mathrm{fm}^{-3}\)) in symmetric nuclear matter. 
As we move from symmetric (\(I_a = 0\)) to an asymmetric medium with \(I_a = 0.3\), the decrease in the magnitude of the quark condensates as a function of density becomes slightly less at higher baryon densities across all considered temperatures, as shown in Fig. \ref{fig_quark_con}(b). Next,  the total contribution of gluon condensate is evaluated using Eq. (\ref{gluon_con}), which consists of two terms. The first term depends solely on the dilaton field $\chi$, while the second term, known as the mass term, depends on $\sigma$, $\zeta$, and $\delta$  scalar fields.
Fig. \ref{fig_gluon_con} shows the variation of individual terms of $\left\langle \frac{\alpha_s}{\pi} G_{\mu \nu}^a G^{a \mu \nu} \right\rangle_{\rho_B}$, denoted as $GG_0$ (first term), $GG_M$ (mass term), and $GG_T$ (total term) with respect to baryonic density ratio, at a fixed value of temperature.
For both $I_a=0$ and $0.3$, the gluon condensate decreases in a manner similar to the quark condensates as a function of $\rho_B$, but with less change in their magnitude compared to quark condensates, as depicted in Fig. \ref{fig_gluon_con} (a) and (b), respectively. We observed that the impact of temperature on the trend of gluon condensate contrasts with its effect on quark condensates. This variation in temperature behavior can be traced to their underlying field dependence. From Fig. \ref{fig_gluon_con}, we noted that $GG_0$ term gives a dominant contribution compared to the $GG_M$ term because the $GG_0$ term incorporates a $\chi^4$ contribution of the dilaton field, as indicated in Eq.( \ref{gluon_con}). Our study revealss that, at nuclear saturation density, the value of \( \langle q\bar{q} \rangle_{\rho_B} \) drops to nearly 36\% and $\left\langle \frac{\alpha_s}{\pi} G_{\mu \nu}^a G^{a \mu \nu} \right\rangle_{\rho_B}$ experiences a reduction of only 6\% of their vacuum values.
  % Specifically, gluon condensate incorporates a $\chi^4$ contribution, along with the $\sigma$ and $\zeta$ fields, as indicated in Eq. \ref{eq_uu} and \ref{eq_dd}, while quark condensates depend on the $\sigma$ and $\delta$ fields, as referenced in Eq. \ref{gluon_con}. 
In literature, similar results of condensates have been reported by utilizing the linear density approximation, the Dirac-Brueckner approach, and the quark meson coupling model-II \cite{Thomas:2007gx, Saito:1997bt, Li:1994mq, Kumar:2019axp}. In Ref. \cite{Cohen:1991nk}, the authors apply the Hellmann-Feynman theorem to derive quark condensates, leading to a 25-50\% reduction in their value, whereas, gluon condensates, estimated through the trace anomaly, show a decrease to merely 3-6\%. As reported in Ref. \cite{Saito:1997bt}, these condensates are assessed using the QMC-II model, quark condensates diminish linearly at low densities, but this reduction becomes less pronounced at higher densities. For gluon condensates, the values fall to 4-6\% from their vacuum. The observed decreases in the quark and gluon condensates are caused by the partial restoration of chiral symmetry, which is crucial for examining the  in-medium properties of mesons using different theoretical models.
% The partial restoration of chiral symmetry, an important factor when analyzing in-medium mass and decay changes using various theoretical models, is responsible for the observed decreases in the quark and gluon condensates.

\section{QCD sum rules (QCDSR) for $D$ and $D^*$ mesons}
\label{QCDSR}
The present section outlines the formulation of the QCD sum rules \cite{Wang:2015uya, Kumar:2019axp}, which is integrated with the chiral SU(3) model to determine the in-medium properties of $D$ (\(D^+\), \(D^0\)) and \(D^*\) (\(D^{*+}\), \(D^{*0}\)) mesons. QCDSR offers a systematic framework for probing the nonperturbative QCD, relying on the Borel transformation as an essential tool to improve the convergence of the perturbative expansion.
We begin with the two-point correlation function, $\Pi_{\mu\nu}(q)$, which can be separated into components corresponding to the vacuum, a static one-nucleon, and a pion bath \cite{Wang:2015uya, Mishra:2014rha, Chhabra:2017rxz}. Thus, we have

\begin{equation}
	\Pi_{\mu\nu}(q) = \Pi^{(0)}_{\mu\nu}(q) + \frac{\rho_B}{2m_N} T^N_{\mu\nu}(q) + \Pi^{P.B.}_{\mu\nu}(q).
	\label{cf}
\end{equation}

In Eq. (\ref{cf}), $T^N_{\mu\nu}$ represents the forward scattering amplitude, and the last term $\Pi^{P.B.}_{\mu\nu}(q)$, known as the pion bath term, has been utilized in previous studies to explore how temperature influences meson properties \cite{Wang:2015uya, Wang:2011fv}. In this study, we omit the pion bath term and instead address temperature dependence on meson properties through the chiral model, which allows the grand thermodynamic potential to explicitly incorporate temperature effects via scalar and vector meson fields \cite{Mishra:2003se,Kumar:2010gb,Kumar:2024lfa,Kaur:2025kjk}. 
% discussed in \cite{ref4}, correlation function $T^N_{\mu\nu}(q)$ can be related to $D^*N$ scattering $T$ matrix through scattering lengths. 
The unknown parameters $a$, $b$, and $c$ involved in the parametrization of the phenomenological spectral densities are computed through the Borel transformation \cite{Wang:2015uya, Wang:2011fv}, leading to the evaluation of the mass shift of pseudoscalar $D$ and vector $D^*$ mesons, which can be defined as,
\begin{equation}
	\Delta m_{D/D^*} = 2\pi \frac{m_N + m_{D/D^*}}{m_N m_{D/D^*}} \rho_B a_{D/D^*}.
	\label{eq_massshiftDD}
\end{equation}

\begin{figure*}[h]
	\centering
	\includegraphics[width=16.5cm]{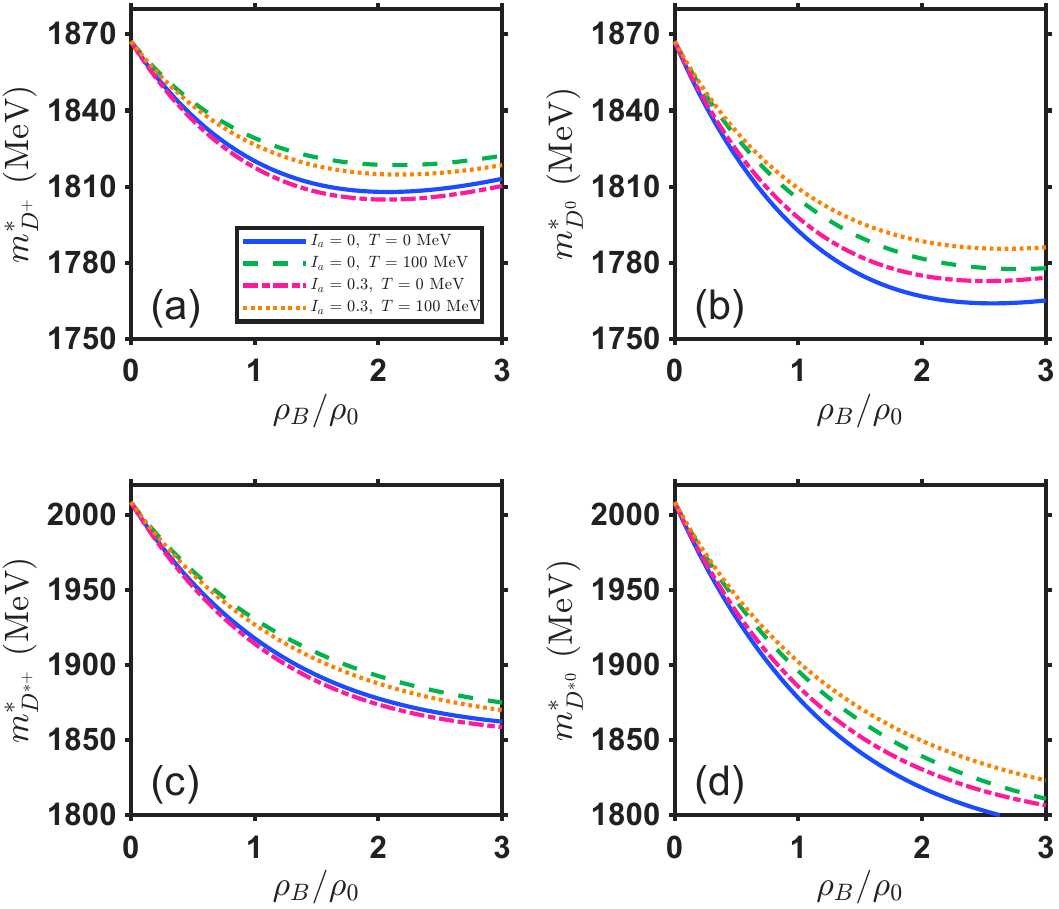}
	\caption{In medium masses of \(D^+\), \(D^0\), \(D^{*+}\), and \(D^{*0}\) mesons as function of the baryon density ratio \( \rho_B/\rho_0 \) within isospin asymmetric nuclear matter,  \( I_a = 0 \) and 0.3, and temperatures, $T=0$ and 100 MeV. }
	\label{fig_massDD}
\end{figure*}

The scattering length for $D/D^*$ mesons, symbolized by $a_{D/D^*}$, is represented as follows \cite{Hayashigaki:2000es}
 \begin{equation}
	a_{D} = \frac{a m_c^2}{f_{D}^2\, m_{D}^4 \left[ -8\pi (m_N + m_{D}) \right]},
	\label{eq_aD}
\end{equation}
and 
\begin{equation}
	a_{D^*} = \frac{a}{f_{D^*}^2\, m_{D^*}^2 \left[ -8\pi (m_N + m_{D^*}) \right]}.
	\label{eq_asD}
\end{equation}
The in-medium masses of $D/D^*$ mesons can be expressed as 
\begin{equation}
	m_{D/D^*}^* = m_{D/D^*} + \Delta m_{D/D^*}.
	\label{eq_massDD}
\end{equation}
The vacuum masses and decay constants of $D$ mesons are specified as $m_{D} = 1.867~\text{GeV}$, $m_{D^{*}} = 2.008~\text{GeV}$, $m_{N} = 0.939~\text{GeV}$, $f_{D} = 0.210~\text{GeV}$, $f_{D^{*}} = 0.270~\text{GeV}$, with the charm quark mass $m_{c} = 1.3~\text{GeV}.$
The work reported in Refs. \cite{Thomas:2007gx, Hilger:2008jg} involves the study of mass splitting of oppositely charged $D$ mesons by dividing the current correlation function into even and odd terms, while this study adopts an approach based on the isospin-average meson current for $D$ and $D^*$ mesons \cite{Hayashigaki:2000es, Wang:2011fv, Chhabra:2016vhp, Kumar:2019axp}. Fig. \ref{fig_massDD} illustrates how the masses of $D$ mesons vary with baryonic density for isospin asymmetry, $I_a = 0$ and $0.3$, and temperatures $T=0$ and 100 MeV. 
% \begin{figure*}[h]
% 	\centering
% 	\includegraphics[width=16cm]{paper_DDs_mass_Ia03_RM.pdf}
% 	\caption{In medium masses of \(D^+\), \(D^0\) and \(D^{*+}\), \(D^{*0}\) mesons as function of the baryon density ratio \( \rho_B/\rho_0 \) within isospin asymmetric resonance matter, with considerations for \( I_a = 0.3 \) at temperatures of $T=0$ and 100 MeV. }
% 	\label{fig_massDD}
% \end{figure*}at nuclear saturation density  (\(\rho_0 0.15\,\mathrm{fm}^{-3}\)). 
The masses of $D$ mesons are evaluated using Eqs.~(\ref{eq_massshiftDD})–(\ref{eq_massDD}), where the parameter $a$ depends on the quark and gluon condensates, which are taken as inputs from the chiral SU(3) model as discussed in detail in Refs.~\cite{Kumar:2019axp, Chhabra:2016uom}. Initially, the Borel window is chosen in a way that ensures the uncertainties in the masses of $D$ ($D^*$) mesons are negligible with respect to the Borel mass parameter $M^2$.
In this work, Borel windows are chosen to be approximately \(4.2-5.3\ \text{GeV}^{3}\) for $D$ mesons and \(4.7-5.6\ \text{GeV}^{3}\) for $D^*$ mesons. We observed that masses of \(D^+\), \(D^0\), \(D^{*+}\), and \(D^{*0}\) mesons decrease as the density of the medium increases at a given value of isospin asymmetry parameter $I_a$ and temperature $T$ as depicted in Figs. \ref{fig_massDD}(a), (b), (c), and (d), respectively. The in-medium values of $m^*_{D}$ at different fixed values of $\rho_B$, $I_a$, and $T$ are tabulated in Table \ref{table_mass}. 
 In Fig. \ref{fig_massDD}(a), with a rise in temperature from zero to 100 MeV, it is noticed that there is less drop in the magnitude of $m^{*}_{D^+}$ as a function of density for both symmetric and asymmetric nuclear medium. We noted that the influence of the isospin asymmetric parameter $I_a$ on the behavior of $m^{*}_{D^+}$ is opposite to its influence on temperature as illustrated in Fig. \ref{fig_massDD}(a). Moreover, within the same environment, the rate of drop in $m^{*}_{D^0}$ is more pronounced compared to $m^{*}_{D^+}$, as one can see in Fig. \ref{fig_massDD}(b). The different masses of the $u$ and $d$ quarks utilized in the calculations of quark condensates [see Eqs. (\ref{eq_uu}) and (\ref{eq_dd})] are responsible for the observed mass difference between $m^{*}_{D^+}$ and $m^{*}_{D^0}$ in symmetric nuclear matter, as one can see from  Table \ref{table_mass}.  
For example, the masses of the \(D^+\), \(D^0\), \(D^{*+}\), and \(D^{*0}\) mesons experience downward shifts of nearly \(47\), \(75\), \(90\), and \(130\) MeV, respectively, from their vacuum values (1867.2~MeV for $D$ and 2008.6~MeV for $D^{*}$) at a temperature of \(T=0\) MeV within isospin symmetric nuclear matter. As the temperature rises to \(T=100\) MeV, these shifts are modified to \(\sim37\), \(62\), \(77\), and \(120\) MeV at nuclear saturation density, respectively.
% The observed mass difference between $m^{*}_{D^+}$ and $m^{*}_{D^0}$ is attributed to the distinct masses of the $u$ and $d$ quarks used in this calculation.
These temperature-dependent changes in the mass shift of $D$ mesons reflect the medium-modified behavior of the quark and gluon condensates. The inclusion of isospin asymmetry into the medium differentiates the behavior of $m^*_{D^+}$ and $m^*_{D^0}$, specifically, for a given value of $\rho_B$, the rate of drop in $m^{*}_{D^+}$ increases and while for $m^{*}_{D^0}$, it decreases as $I_a$ rises as one can see in Fig. \ref{fig_massDD}(a) and (b).     
The mass of vector $D^*$ mesons decreases more significantly with an increase in baryonic density ratio at a specific temperature than that of pseudoscalar $D$ mesons, which is attributed to the stronger attractive interaction of the $D^*$meson with the nuclear medium, as illustrated in Fig. \ref{fig_massDD}(c) and (d). 
In literature, the QCD sum rules method has been utilized in various studies to investigate the in-medium modifications of $D$ meson masses at nuclear saturation density $\rho_0$. For instance, in Ref. \cite{Wang:2015uya}, the authors incorporated both leading-order and next-to-leading-order corrections in symmetric nuclear matter at zero temperature. The NLO analysis predicts a mass shift of $D$ mesons to be $-72$ MeV, while considering only the LO terms results in a smaller shift of $-47$ MeV in $D$ meson mass at $\rho_0$. Also, the study presented in Ref. \cite{Hilger:2008jg}, the Lehmann representation for the correlation function was employed, and the QCDSR was divided into odd and even components, leading to a splitting of approximately 60 MeV between $D^{+}$ and $D^{-}$ mesons. According to Ref. \cite{Tsushima:1998ru}, the mass shifts of $D^{+}$ and $D^{-}$ mesons were determined to be around $-62$ MeV at nuclear matter density within the quark-meson coupling model. The in-medium characteristics of pseudoscalar $D$ mesons have been analyzed within a nuclear environment using an effective chiral model, which explicitly includes $D$ meson-nucleon interactions via the dispersion relation. At $\rho_0$, the mass shift for $D^+$ and $D^0$ mesons is estimated to lie between -72 and -62 MeV \cite{Kumar:2010gb}. Moreover, Ref.~\cite{Jimenez-Tejero:2011dif} forecasts opposite mass shift for the $D$ mesons, with nearly $+32$ for  $D^{-}$ and $-27$ MeV for $D^{+}$, at $\rho_0 = 0.17~\mathrm{fm}^{-3}$ using a self-consistent coupled-channel approach.
% Ref. \cite{Tsushima:1998ru} reports that the mass shifts of  $D^{+}$ and $D^{-}$ mesons were found to approximately $-62$ MeV at nuclear matter density by employing the quark- meson couping model. Using an effective chiral model, the in-medium properties of pseudoscalar $D$ mesons have been examined within a nuclear medium, incorporates explicittly $D$ meson-nucleon interactions via the dispersion relation. At $\rho_0$, the mass shift for $D$ mesons ($D^+, D^0$) is estimated to range from 72 to 62 MeV \cite{Kumar:2010gb}. Furthermore, in Ref.~\cite{Jimenez-Tejero:2011dif},  the authors predicted mass shift of the $D^{-}$ ($D^{+}$) meson of around $+32$ ($-27$) MeV, respectively, at$\rho_0 = 0.17~\mathrm{fm}^{-3}$ within a self-consistent coupled-channel approach
% These previous studies focused exclusively on meson properties in the nuclear medium.
\begin{table}[h!]
	\centering
    \caption{In-medium values of $m^*_{D}$ are presented at specific constant values of $\rho_B$, $I_a$, and $T$ (in MeV).}
	
	\renewcommand{\arraystretch}{1.5} % <-- increases vertical spacing
	\begin{tabular}{|c|c|c|c|c|c|c|c|c|}
		\hline
		& \multicolumn{4}{c|}{T=0 MeV}  & \multicolumn{4}{c|}{T=100 MeV}  \\
		\cline{2-9}
		$m^*_{D}$ (MeV) & \multicolumn{2}{c|}{$I_a=0$} & \multicolumn{2}{c|}{$I_a=0.3$} & \multicolumn{2}{c|}{$I_a=0$} & \multicolumn{2}{c|}{$I_a=0.3$} \\
		\cline{2-9}
		& $\rho_0$ & $3\rho_0$ & $\rho_0$ & $3\rho_0$ & $\rho_0$ & $3\rho_0$ & $\rho_0$ & $3\rho_0$ \\ \hline
		
		${D^+}$ & ~1820~ & ~1813~ & ~1817~ & ~1810~ & ~1830~ & ~1822~ & ~1826~ & ~1818~ \\ 
\hline
${D^0}$ & ~1792~ & ~1765~ & ~1797~ & ~1774~ & ~1805~ & ~1777~ & ~1809~ & ~1786~ \\
\hline
${D^{*+}}$ & ~1918~ & ~1862~ & ~1914~ & ~1858~ & ~1931~ & ~1875~ & ~1927~ & ~1870~ \\
\hline
${D^{*0}}$ & ~1878~ & ~1793~ & ~1885~ & ~1806~ & ~1896~ & ~1811~ & ~1902~ & ~1823~ \\

		\hline
		
	\end{tabular}
	
	\label{table_mass}
\end{table}

\section{The effective Lagrangian approach}
\label{ELA}
% The $J/\psi$ mass shift within the medium results from the modified contributions of the $DD$, $DD^*$, and $D^*D^*$ meson loops to its self-energy, computed using an effective Lagrangian method under the flavor SU(4) symmetry, incorporating quark meson coupling model as detailed in Refs. \cite{Zeminiani:2020aho, Krein:2010vp, Krein:2017usp, Tsushima:2011fg}. In this study, we predict these results for isospin asymmetric nuclear matter at zero and finite temperatures by employing an effective Lagrangian in combination with a QCD sum-rule method.
In this section, we predict the results of the mass shift of the $J/\psi$ meson in isospin asymmetric nuclear matter, utilizing an effective Lagrangian approach under the flavor SU(4) symmetry, combined with QCD sum rules. We account for $DD$ and $DD^*$ meson loop contributions to $J/ \psi$  self-energy.
% Previous studies, in Refs. \cite{Zeminiani:2020aho, Krein:2010vp, Krein:2017usp, Tsushima:2011fg}, utilized quark-meson coupling model to calculate in-medium properties of the $J/\psi$ meson within symmetric nuclear matter. 
% at temperature $T=0$.
% This study investigates the $J/\psi$ meson mass shift in isospin-asymmetric nuclear matter. By employing an effective Lagrangian approach under $SU(4)$ flavor symmetry coupled with QCD sum rules, we account for $DD$ and $DD^*$ meson loop contributions to the $J/ \psi$ self-energy. While previous works \cite{Zeminiani:2020aho, Krein:2010vp, Krein:2017usp, Tsushima:2011fg} utilized quark-meson coupling models for symmetric matter at $T=0$, this work extends the analysis to asymmetric environments
% We predict the in-medium mass shift of the $J/\psi$ meson by evaluating self-energy corrections arising from $DD$ and $DD^*$ meson loops. Our framework integrates an effective Lagrangian based on $SU(4)$ symmetry with the QCD sum-rule approach. In contrast to earlier investigations \cite{Zeminiani:2020aho, Krein:2010vp, Krein:2017usp, Tsushima:2011fg}, which focused on isospin-symmetric matter via quark-meson coupling, our investigation addresses the specific influence of isospin asymmetry.
We utilize the effective Lagrangian densities to describe the interaction of $J/\psi$ with $D$ and ${D}^*$ mesons (the field corresponding to $J/\psi$ will be denoted as $\psi$) \cite{Lin:1999ad, Haidenbauer:2007jq, Krein:2010vp}, and given as
\begin{equation}
	\mathcal{L}_{\psi DD} = i g_{\psi DD} \psi^{\mu} 
	\left[ \bar{D} (\partial_{\mu} D) - (\partial_{\mu} \bar{D}) D \right],
\end{equation}

\begin{equation}
	\mathcal{L}_{\psi DD^*} = \frac{g_{\psi DD^*}}{m_{\psi}} 
	\varepsilon_{\alpha \beta \mu \nu}
	(\partial^{\alpha} \psi^{\beta})
	\left[ (\partial^{\mu} \bar{D}^{*\nu}) D + \bar{D} (\partial^{\mu} D^{*\nu}) \right],
\end{equation}

\begin{equation}
	\begin{split}
		\mathcal{L}_{\psi D^* D^*} = i g_{\psi D^* D^*} \big[ 
		& \psi^{\mu} \left( (\partial_{\mu} \bar{D}^{*\nu}) D^*_{\nu} 
		- \bar{D}^{*\nu} (\partial_{\mu} D^*_{\nu}) \right) \\
		& + (\partial_{\mu} \psi^{\nu}) \bar{D}^*_{\nu} D^{*\mu} 
		- \psi^{\nu} (\partial_{\mu} \bar{D}^*_{\nu}) D^{*\mu} \\
		& + \bar D^{*\mu} \left( \psi^{\nu} (\partial_{\mu} {D}^*_{\nu}) 
		- (\partial_{\mu} \psi^{\nu}) {D}^*_{\nu} \right) \big].
	\end{split}
\end{equation}
The following convention is employed for the isospin doublets of the pseudoscalar and vector $D$-mesons:
\begin{equation}
	D = 
	\begin{pmatrix}
		D^{+} \\
		D^{0}
	\end{pmatrix},
	\quad
	\bar{D} = 
	\begin{pmatrix}
		D^{-} & \bar{D}^{0}
	\end{pmatrix},
	\quad
	D^{*} = 
	\begin{pmatrix}
		D^{*+} \\
		D^{*0}
	\end{pmatrix},
	\quad
	\bar{D}^{*} = 
	\begin{pmatrix}
		D^{*-} & \bar{D}^{*0}
	\end{pmatrix}.
\end{equation}
If SU(4) symmetry treated as exact, it would establish the identical relationships among the coupling constants in the Lagrangian, which are taken to be 
$g_{\psi DD} = g_{\psi D D^*} = g_{\psi D^*D^*} = 7.64$.
These coupling constants are determined by using the vector-meson dominance model \cite{Sakurai:1960ju, Lin:2000ke}.
%We are focused on the difference between the in-medium and the vacuum mass of $J/\psi$, also referred to as the Lorentz scalar potential.  
The mass shift of $J/\psi$ meson is given as  
\begin{equation}
	\Delta m_{\psi} = m_{\psi}^* - m_{\psi},
	\label{eq_massshift_J}
\end{equation}
with in-medium masses of $J/\psi$ meson derived from
\begin{equation}
	m_{\psi}^{*2} = (m_{\psi}^0)^2 + \Sigma_\ell (k^2 = m_{\psi}^{*2}).
\end{equation}
The term $\Sigma_\ell (k^2)$ is the total 
$J/\psi$ self-energy from different loop contributions with $\ell$ represents $DD, DD^*, D^*D^*$ meson loops and $m_{\psi}^0$ stands for the bare mass. 
By taking the average over the three $J/\psi$ polarization, $J/\psi$ self-energy $\Sigma_{\ell}(k^2)$, can be expressed as follows \cite{Krein:2010vp, Zeminiani:2020aho} 
\begin{equation}
	\Sigma_{\ell}(m_{\psi}^{*2}) = -\frac{g_{\psi \ell}^2}{3 \pi^2}
	\int dq\, \textbf{q}^2\, F_{\ell}^2(\textbf{q}^2)\, K_{\ell}(\textbf{q}^2),
\end{equation}

where $K_{\ell}(\textbf{q}^2)$ for each loop contribution are given by

\begin{equation}
    K_{DD}(\textbf{q}^2) = \frac{\textbf{q}^2}{\omega^*_D} \left( \frac{1}{\omega_D^{*2} - m_{\psi}^{*2}/4} \right),
\end{equation}

\begin{equation}
	K_{DD^*}(\textbf{q}^2) = \frac{\textbf{q}^2}{\omega^{*}_D \omega^{*}_{D^*}}
	\frac{\bar {\omega^{*}}_D }{\bar{\omega^{*}}_D^2 - m_{\psi}^{*2} / 4},
\end{equation}

\begin{equation}
	K_{D^*D^*}(\textbf{q}^2) = \frac{1}{4 m_{\psi} \omega^{*}_{D^*}}
	\left[ \frac{A(q^0 = \omega^{*}_{D^*})}{\omega^{*}_{D^*} - m_{\psi}^{*}/2}
	+ \frac{A(q^0= \omega^{*}_{D^*}-m_{\psi}^*)}{\omega^{*}_{D^*} + m_{\psi}^{*}/2} \right],
\end{equation}
here $\omega^{*}_D = (\textbf{q}^2 + m_D^{*2})^{1/2}$, 
$\omega^{*}_{D^*} = (\textbf{q}^2 + m_{D^*}^{*})^{1/2}$, 
$\bar{\omega^{*}}_D = (\omega^{*}_D + \omega^{*}_{D^*})/2$, and $m^*_{D/D^*}$ represents the isospin-averaged mass of the $D/D^*$ meson, calculated as the average of the charged and neutral states $m^*_{D^+/D^{*+}} + m^*_{D^0/D^{*0}} $.
In above expression , $A(q) = \sum_{i=1}^{4} A_i(q)$, with

\begin{align*}
	A_1(q) &= -4q^2 \left[ 4 - \frac{q^2 + (q - k)^2}{m_{D^*}^{*2}}
	+ \frac{[q \cdot (q - k)]^2}{m_{D^*}^{*4}} \right], \\[6pt]
	A_2(q) &= 8 \left[ q^2 - \frac{[q \cdot (q - k)]^2}{m_{D^*}^{*2}} \right]
	\left[ 2 + \frac{(q^0)^2}{m_{D^*}^{*2}} \right], \\[6pt]
	A_3(q) &= 8 (2q^0 - m^{*}_\psi) \left[ q^0 - (2q^0 - m^{*}_\psi) \frac{q^2 + q \cdot (q - k)}{m_{D^*}^{*2}}
	+ q^0 \frac{[q \cdot (q - k)]^2}{m_{D^*}^{*4}} \right], \\[6pt]
	A_4(q) &= -8 \left[ q^0 - (q^0 - m^{*}_\psi) \frac{q \cdot (q - k)}{m_{D^*}^{*2}} \right]
	\left[ (q^0 - m^{*}_\psi) - q^0 \frac{q \cdot (q - k)}{m_{D^*}^{*2}} \right].
\end{align*}
In the above expressions, the four-vectors $q$ and $k$ are defined as $q = (q^0, \mathbf{q})$ and $k = (m^{*}_{\psi}, \mathbf{0})$. To regulate the self-energy loop integrals, we apply phenomenological form factors $F_\ell(\textbf{q}^2)$. The $F_\ell(\textbf{q}^2)$ signifies the product of the vertex form factors $\big(u_\ell(\textbf{q}^2)\big)$, which is applied to both vertices, given as~\cite{Tsushima:2011kh,Leinweber:2001ac, Tsushima:2018goq, Lin:1999ad}.
\begin{equation}
    u_{D,D^*}(\textbf{q}^2) = 
    \left(
    \frac{\Lambda_{D}^2 + m_{\psi}^{*2}}
    {\Lambda_{D}^2 + 4\omega_{D/D^*}^{*2}(\mathbf{q^2})}
    \right)^2.
    \label{form_factor}
\end{equation}
The dipole form factors are given by
$F_{DD}(\mathbf{q}^2) = u_D^2(\mathbf{q}^2)$,
$F_{DD^*}(\mathbf{q}^2) = u_D(\mathbf{q}^2)u_{D^*}(\mathbf{q}^2)$, and
$F_{D^*D^*}(\mathbf{q}^2) = u_{D^*}^2(\mathbf{q}^2)$, where, $\Lambda_D$ is the cut-off mass parameter. The choice of $\Lambda_D$ for the interaction of $J/\psi $ meson with $DD, DD^*, D^*D^*$ meson loops plays an important role, as they significantly influence the calculated results. These form factors are introduced to represent the finite spatial extent of the mesons and their overlapping region at the vertices. When the cutoff parameters are comparable to the meson masses, their associated Compton wavelengths are of the same order as the sizes of the mesons. This range of cutoff parameter values reduces the physical significance of the form factors \cite{Zeminiani:2020aho}. Consequently, it is crucial to select cutoff values with care to accurately represent the finite-size effects of the interacting mesons. We choose $\Lambda_D = 1$ to 3~\text{GeV} to calculate the in-medium mass of $J/\psi$ using heavy quark and meson symmetry in QCD, which is consistent with the parameters used in the study reported in Refs. \cite{Zeminiani:2020aho, Krein:2010vp}.
\begin{figure*}[h]
	\centering
	\includegraphics[width=16cm]{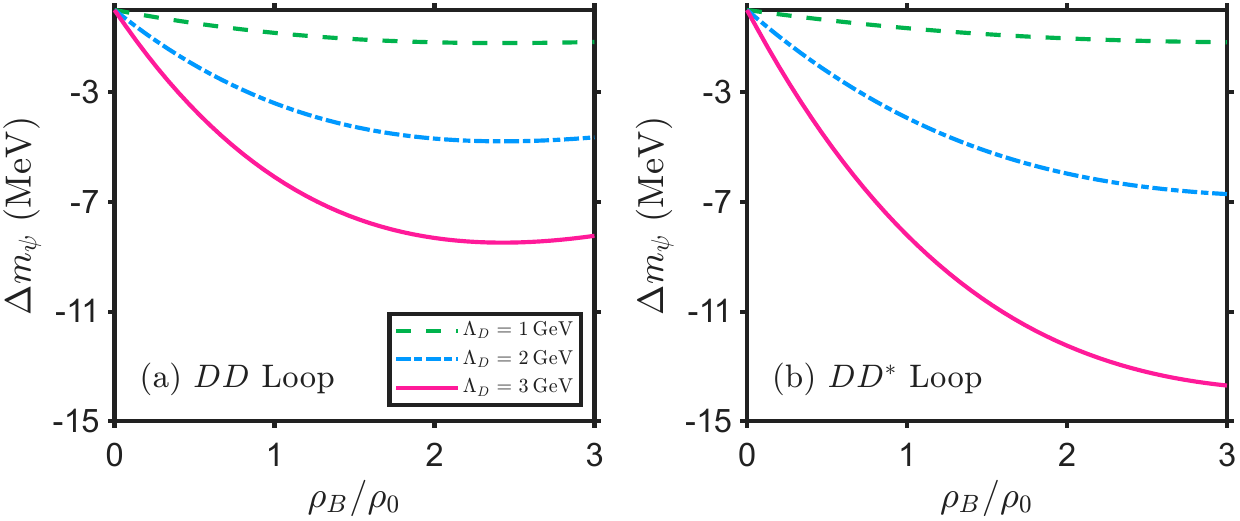}
	\caption{Density dependence of the $DD$ and $DD^*$ loop contribution to the $J/\psi$ mass shift ($\Delta m_{\psi}$) in asymmetric nuclear matter at $I = 0.3$ and $T=100$ MeV for different values of
		the cutoff $\Lambda_{D}$.
	}
	\label{FIG_MS_JDDs1}
\end{figure*} 
\begin{figure*}[h]
	\centering
	\includegraphics[width=16cm]{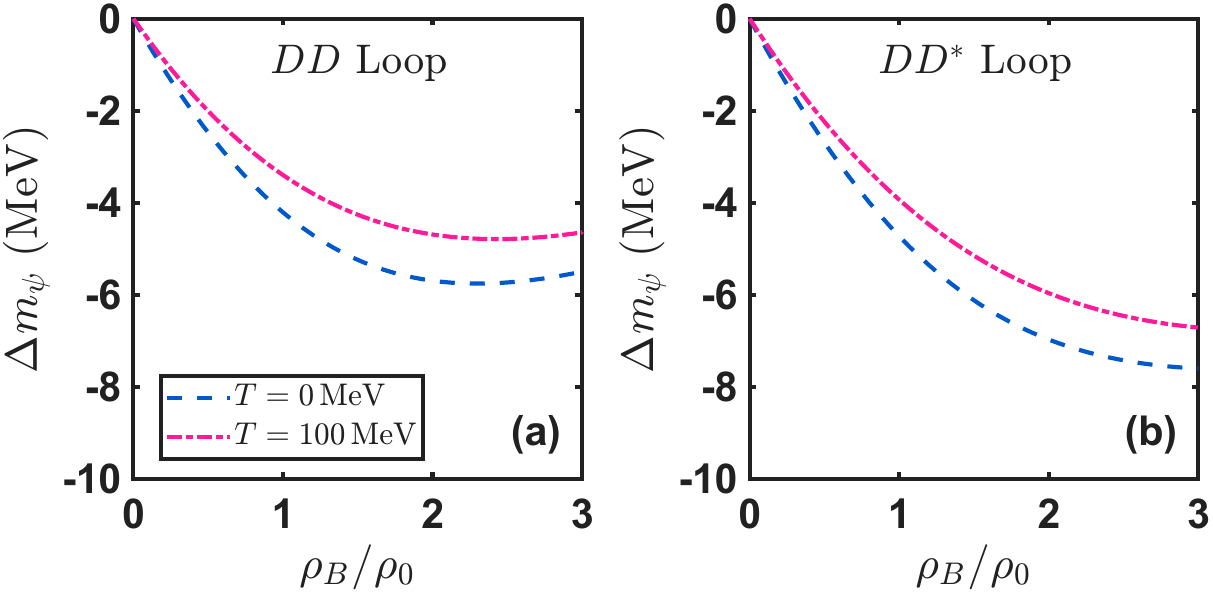}
	\caption{Density dependence of the $DD$ and $DD^*$ loop contributions to the $J/\psi$ mass shift ($\Delta m_{\psi}$) in asymmetric nuclear matter at $I = 0.3$ and $\Lambda_{D}=2$ GeV for different values of temperatures.}
	\label{FIG_MS_JTTDDs2}
\end{figure*} 
\begin{figure*}[h]
	\centering
	\includegraphics[width=9cm]{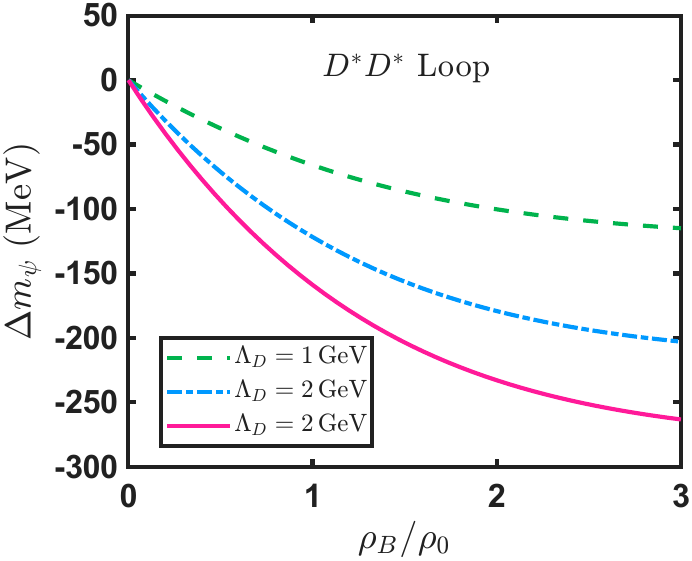}
	\caption{The $D^*D^*$ loop contribution to the $J/\psi$ mass shift as function of the baryon density ratio \( \rho_B/\rho_0 \)  is examined in symmetric nuclear matter at $T=0$ MeV, with various cutoff values $\Lambda_{D}$.}
	\label{FIG_MS_JDsDs}
\end{figure*} 

Next, we proceed to present the results for the in-medium mass shift of the $J/\psi$ meson, $\Delta m_{J/\psi}$, which are calculated from Eq. (\ref{eq_massshift_J}). To begin, we analyze the individual loop contributions of $\psi DD$, $\psi DD^*$, and $\psi D^*D^*$ separately, focusing on their impact on the mass shift of $J/\psi$ in relation to baryonic density. This analysis is conducted for various cutoff mass values $\Lambda_D$ and temperature conditions. In Fig. \ref{FIG_MS_JDDs1}(a), the findings of the in-medium mass shift of the $J/\psi$ meson for the $DD$ loop are depicted as a function of the $\rho_B/\rho_0$ ratio in nuclear matter, with $I_a=0.3$ and $T=100$ MeV, using a cutoff mass parameter ranging from 1 to 3 \text GeV. We observe that as the density of the medium increases, the magnitude of $\Delta m_{J/\psi}$ becomes more negative for a given $\Lambda_D$. 
% This negative mass shift implies that $J/\psi$ is attracted to nuclear mean fields.
When moving to higher $\Lambda_D$ values, a more substantial downward shift in $\Delta m_{J/\psi}$ is observed. 
% In other words, $\Lambda_D$ approaches the vacuum mass of $J/\psi$, and the mass shift diminishes.
From the $DD^*$ loop contribution, a more significant negative mass shift of the $J/\psi$ meson is noticed, as depicted in Fig. \ref{FIG_MS_JDDs1}(b), which is consistent with the findings reported in Ref. \cite{Zeminiani:2020aho}. This negative mass shift of $J/\psi$ indicates the strong coupling of vector $D^*$ mesons with the scalar and vector fields of the medium compared to the pseudoscalar $D$ meson. 
% Consequently, the vector interactions are stronger for $D^*$ mesons compared to $D$ mesons, and since vector mesons are heavier, their loop propagators are more significantly suppressed. 
As a result, there is a greater drop for the $DD^*$ meson loop contribution compared to the $DD$ meson loop. To explicitly examine the temperature dependence, the mass shift of $J/\psi$ meson has been evaluated at temperature $T=0$ and 100 MeV while keeping $\Lambda_D$ fixed at 2~GeV 
as illustrated in Fig. \ref{FIG_MS_JTTDDs2}. The drop in magnitude of $\Delta m_{J/\psi}$ at $T=100$ MeV is less pronounced than that at $T=0$ MeV for both $DD$ and $DD^*$loops, as one can see from Fig. \ref{FIG_MS_JTTDDs2}(a) and (b). We listed the prediction of mass shift of $J/\psi$ with the dependence of $DD$ and $DD^{*}$ meson-loop contribution at \(\rho_0 = 0.15\,\mathrm{fm}^{-3}\) within nuclear medium in Table \ref{table_jpsi}. Our analysis indicates a negative mass shift of the $J/\psi$ within an isospin-asymmetric nuclear medium, varying from approximately $-1.5$ to $-14$ MeV at nuclear saturation density. This variation arises from the choice of cutoff mass parameter in the range $\Lambda_D=1$ to 3 GeV and shows good consistency with the most recent theoretical analyses of the $J/\psi$ meson \cite{Cobos-Martinez:2025iqg}, where the mass shift of the $J/\psi$ is estimated to be between $-4$ and $-12$ MeV within same environment by using QMC Model, attributed solely to the $DD$ loop contribution \cite{CobosMartinez:2021bia}.
In Ref. \cite{Chhabra:2020dsr}, the authors examined the mass shift of $J/\psi$ within a symmetric nuclear medium and a strange medium, focusing only on the contributions from the $DD$ meson loop.
% The medium effects were incorporated through a combined approach utilizing the hadronic chiral SU(3) model and QCD sum rules. 
The study determined that the mass shift of $J/\psi$ is -3.5~MeV in cold nuclear matter and -4.5~MeV in strange matter, respectively.
% In Ref. \ref{}, the authors conducted an investigation into the mass shift of $J/\psi$ within both symmetric nuclear and strange media, attributing the effects solely to the $D$ meson loop contributions. The study employed a combined methodology integrating the hadronic chiral SU(3) model with QCD sum rules to account for medium effects. The findings revealed that the mass shift of $J/\psi$ is -3.5 MeV in cold nuclear matter and -4.5 MeV in strange matter, respectively.
\begin{table}[h!]
	\centering
    \caption{Dependence of the $J/\psi$ mass shift on cutoff parameter, $\Lambda_D$, through the contributions of  $DD$ and $DD^*$ meson loops at \(\rho_0 = 0.15\,\mathrm{fm}^{-3}\) within an isospin asymmetric nuclear medium.}
	
	\begin{tabular}{|c|c|c|c|}
	\hline 
	$\Lambda_D$ (GeV) ~& $\Delta m^{DD}_{J/\psi}$ (MeV) ~& $\Delta m^{DD^{*}}_{J/\psi}$ (MeV) ~& $\Delta m^{DD+DD^{*}}_{J/\psi}$ (MeV) \\ \hline
	1 & -0.83 & -0.66 & -1.49 \\ \hline
	2 & -3.40 & -3.93 & -7.32 \\ \hline
	3 & -6.08 & -8.20 & -14.28 \\ \hline 
\end{tabular}
	
	\label{table_jpsi}
\end{table}
In Fig. \ref{FIG_MS_JDsDs}, we present the results of the $D^*D^*$ meson loop contribution to $J/\psi$ mass shift. An unexceptional large downward mass shift of $J/\psi$ within an isospin symmetric nuclear medium is observed, ranging from $-70$ to $-158$~MeV at nuclear saturation density and depending on the value of the cutoff mass parameter. Our results are in close agreement with those reported in Ref. \cite{Zeminiani:2020aho}, where masses of $J/\psi$ for $D^*D^*$ meson loop were found to lie between -61 to -164 MeV using the quark meson coupling model. However, from a basic quantum mechanical perspective, it is anticipated that heavier loops would have a lesser impact. To ensure the accuracy of our minimum prediction and to avoid possible model-dependent uncertainties, we concentrate on the $DD$ and $DD^{*}$ meson-loop contribution \cite{Zeminiani:2020aho}. These findings may offer significant insights into the mechanism responsible for $J/\psi$ production and suppression, as observed in various experiments, such as the NA38, and NA60 collaborations at CERN SPS  \cite{NA38:1989yxm, NA38:1994yau, NA60:2007odv}, the PHENIX experiments at RHIC \cite{PHENIX:2006gsi, PHENIX:2011img}, and the ALICE experiment at the LHC \cite{ALICE:2012jsl}.
\section{$J/\psi$-NUCLEI BOUND STATE}
\label{sec_bound_state}
In this section, we predict the existence of bound states of the $J/\psi$ meson with nuclei $\text{O}^{16}$, $\text{Ca}^{40}$, $\text{Zr}^{90}$, and $\text{Pb}^{208}$ by calculating their binding energy and absorption decay width. In a finite nucleus, the scalar quark and gluon condensates are evaluated as a function of the radial coordinate $r$, within the range $0 \leq r \leq R$, where $R$ denotes the cutoff radius beyond which the baryonic density is considered zero \cite{DeJager:1974liz}. To determine their radial dependence, we solve the coupled equations of motion for mesonic fields, which are associated with baryonic densities within the chiral SU(3) model.
The total baryonic density $\rho_B(r)$ of nuclei is $\rho_B(r) = \sum\limits_{i=p,n} \rho_i(r)$ \cite{Nieves:1993ev, Kumar:2024lfa}.
% In a finite nucleus, the scalar quark and gluon condensates are evaluated as a function of the radial coordinate $r$, within the range $0 \leq r \leq R$, where $R$ denotes the cutoff radius beyond which the baryonic density is considered zero.
For nuclei with atomic number $A<18$, the harmonic potential type density distribution is used and is represented as follows \cite{DeJager:1974liz, Nieves:1993ev}  
\begin{equation}
\rho_i(r) = \rho_{i,0} \left( 1 + a_i \left( \frac{r}{R_i} \right)^2 \right) 
\exp \left[ - \left( \frac{r}{R_i} \right)^2 \right],
\end{equation}
while the two-parameter Fermi distribution density is utilized for heavy nuclei, given as
 \begin{equation}
\rho_i(r) = \frac{\rho_{i,0}}{1 + \exp \left[ \frac{r - R_i}{a_i} \right]}.
\end{equation}
 
 The center density of nucleons in the above equations is represented by $\rho_{i,0}$, while the associated radius and diffuseness parameters are denoted by $R_i$ and $a_i$, respectively (for nucleon of type $i$), and expressed as
 \begin{align}
R_i &= R_{i,e} + \frac{5 r_i^2 R_{i,e}}{15 R_{i,e}^2 + 7 \pi^2 a_{i,e}^2}, 
\end{align}
and
\begin{align}
a_i &= \left( \frac{R_{i,e}^3 + \pi^2 a_{i,e}^2 R_{i,e} - R_{i,e}^3}{\pi^2 R_{i,e}} \right)^{1/2},
\end{align}
 here $r^{2}_{i}=0.69$~$\text{fm}^{2}$, represents mean-squared radius of nucleons. Table \ref{table_Ra} presents the nucleon density parameter $R_{i,e}$ and $a_{i,e}$ for different nuclei \cite{Nieves:1993ev}.
 Isospin asymmetry is crucial in understanding the properties of strongly interacting matter, particularly in neutron-rich heavy nuclei. The global asymmetry parameter is expressed as $I_a = (N - Z)/A$, where $N$ represents the neutron number, $A$ is the mass number, and $Z$ denotes the proton number. While, the neutron-to-proton asymmetry is defined as $I_a = \frac{\rho_n(r) - \rho_p(r)}{2\rho_B(r)}$, which tends to reach unity at the nuclear surface \cite{Steiner:2004fi, Steiner:2005jma}.
% While the global asymmetry is defined by $I = (N - Z)/A$, the local neutron-to-proton asymmetry $\delta = (\rho_n - \rho_p)/\rho$ approaches unity at the nuclear periphery. The fundamental properties of nuclear systems—ranging from binding energies and density distributions to collective excitations—are governed by the isospin dependence of the strong interaction, typically quantified by a symmetry energy of $E_{sym} \approx -30$ MeV.
 % The isospin asymmetry parameter is defined as $I = (N - Z)/A$, where $ N$ is the neutron number, $ A$ is the mass number, and $ Z$ is the proton number. It can be represented in terms of nucleon densities as $I = \frac{\rho_n(r) - \rho_p(r)}{2\rho_B(r)}$.
\begin{table}[h!]
\centering

\caption{Nucleons density parameters $R_{p,e}$, $R_{n,e}$, $a_{p,e}$ and $a_{n,e}$ for different nuclei $\text{O}^{16}$, $\text{Ca}^{40}$, $\text{Zr}^{90}$, and $\text{Pb}^{208}$ (in $\text{fm}$).}
\begin{tabular}{|c|c|c|c|c|}
\hline 
 & \multicolumn{2}{c|}{Proton} & \multicolumn{2}{c|}{Neutron} \\ \cline{2-5} 
Nucleus & $R_{p,e}$ (fm) & $a_{p,e}$ (fm) & $R_{n,e}$ (fm) & $a_{n,e}$ (fm) \\ \hline
$\text{O}^{16}$ & 1.833 & 1.544 & 1.815 & 1.529 \\ \hline
$\text{Ca}^{40}$ & 3.51 & 0.563 & 3.43 & 0.563 \\ \hline
$\text{Zr}^{90}$ & 4.484 & 0.55 & 4.94 & 0.55 \\ \hline
$\text{Pb}^{208}$ & 6.624 & 0.549 & 6.89 & 0.549 \\  \hline
\end{tabular}
\label{table_Ra}
\end{table}

At relatively low momenta, the meson experiences an effective potential within the nucleus, which serves as input to the Klein-Gordon equation to investigate the bound states of mesic-nuclei. The Klein-Gordon equation, incorporating a local complex potential $V(r)$, can be expressed as follows \cite{Cobos-Martinez:2023hbp}   
\begin{equation}
\label{eq_KG}
\left( -\nabla^2 + (\mu + V(r))^2 \right)\psi(r) = \epsilon^2 \psi(r),
\end{equation}
here, the parameter $\mu$ denotes the reduced mass of the mesic-nuclei system and $\epsilon$ stands for the energy eigenvalue of Eq. (\ref{eq_KG}). For a local complex potential, $V(r)$, we write \cite{Metag:2017yuh, Tsushima:1998qp} 
\begin{equation}
V(r) = U(r) - \frac{i}{2} W(r).
\end{equation}
 The real part of the potential  $U(r)$ reflects the mass modification of the meson, whereas the imaginary part  $W(r)$ accounts for their absorption within the nucleus. These are defined through the relations,  
 \begin{figure*}[h]
	\centering
	\includegraphics[width=14cm]{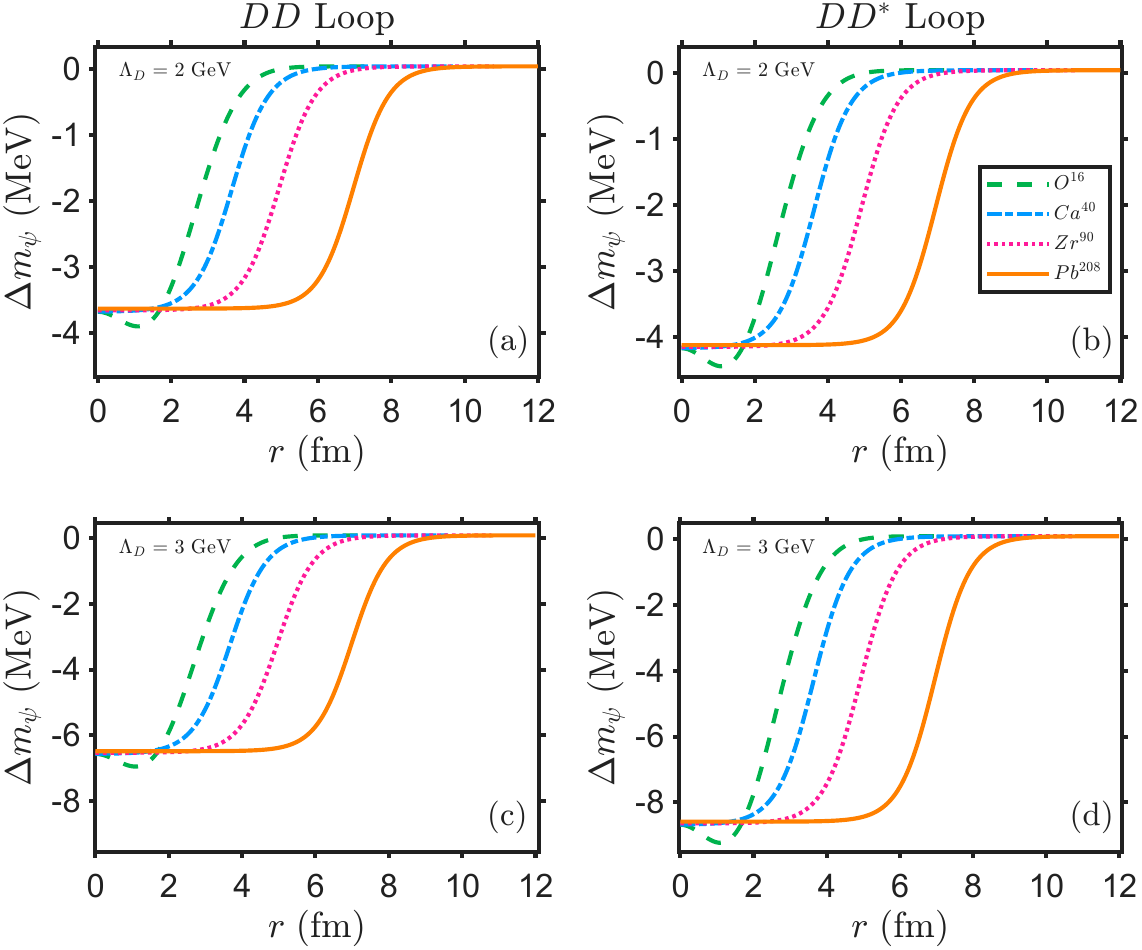}
	\caption{The result of $J/\psi$ mass shift as a function of $r$ in nuclei $\text{O}^{16}$, $\text{Ca}^{40}$, $\text{Zr}^{90}$, and $\text{Pb}^{208}$ at different value of cut off parameter $\Lambda_D$, from the $DD$ loop contribution, illustrated in subfigures (a) and (c), and the $D D^*$ loop contribution, presented in subfigures (b) and (d).}
	\label{FIG_r_diff}
\end{figure*}
\begin{equation}
\label{eq_real}
U(r) = m_{\psi}^*(r) - m_{\psi} = \Delta m_{\psi}(\rho_{i,0})\frac{\rho_B(r)}{\rho_{i,0}},
\end{equation}
and
% \begin{equation}
% W(r) = \Gamma_0(\rho_{i,0})\frac{\rho_B(r)}{\rho_{i,0}},
% \end{equation}
\begin{equation}
\label{eq_img}
W(r) = \left( -\kappa\, \Delta m_{\psi}(\rho_{i,0}) + \Gamma_{\text{vac}} \right)\,\frac{\rho_B(r)}{\rho_{i,0}},
\end{equation}
respectively. In the above equation, 
$\Delta m_{\psi}(\rho_{i,0})$ is the mass shift of the $J/\psi$ meson at the central baryonic density of nuclei, $\kappa$ mimics the absorption effect, and $\Gamma_{vac}$ is the vacuum decay width, which is not taken in the calculation. 
% To calculate $  \Gamma_{0}(\rho_{i,0})=( -\kappa\, \Delta m_{\psi}(\rho_{i,0}) + \Gamma_{\text{vac}})$, we use the same method that was adopted in caculation of the QMC model.
Following this, we calculate the binding energy and absorption decay width of the $J/\psi$ meson by solving Eq. (\ref{eq_KG}) in momentum space. This calculation is performed using a Fourier transform, which incorporates partial wave decomposition of the potential \cite{Landau:1982iu, Kwan:1978zh}. This method enables us to derive complex eigenvalues, where the real part signifies the binding energy $E$, and the imaginary part indicates the decay width $\Gamma$, as defined by the equations $E=\text{Re}~\epsilon-\mu$ and $\Gamma = -2\text{Im}~\epsilon$, respectively.
%DOI: 10.1103/PhysRevC.109.025202
\begin{figure*}[h]
	\centering
	\includegraphics[width=14cm]{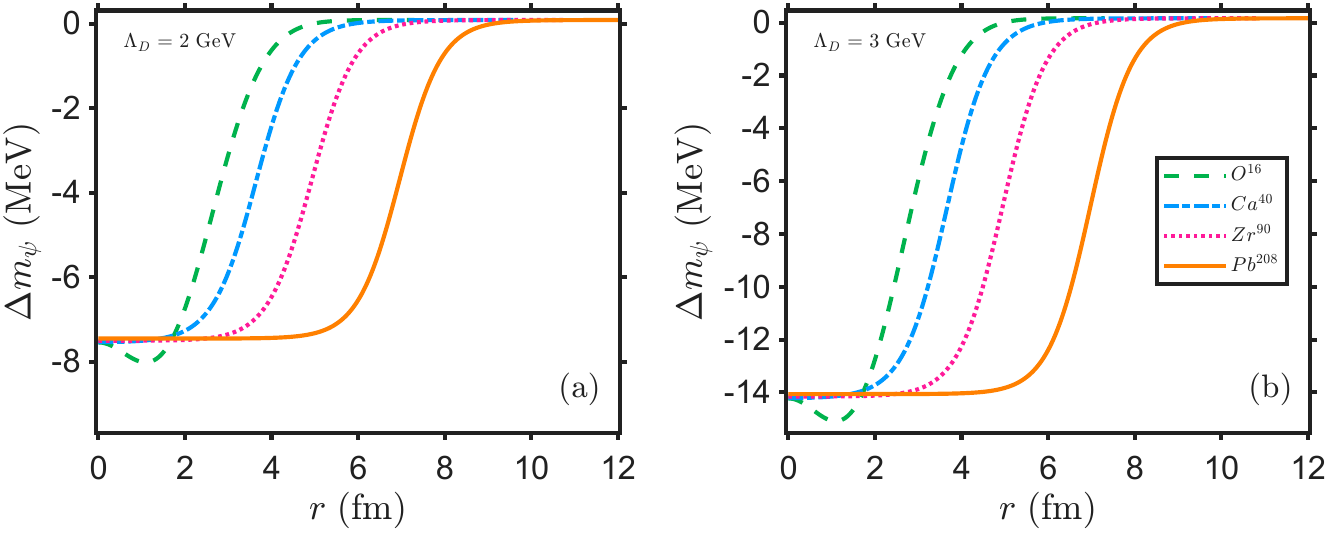}
	 \caption{The $J/\psi$ mass shift as a function of $r$ in nuclei $\text{O}^{16}$, $\text{Ca}^{40}$, $\text{Zr}^{90}$, and $\text{Pb}^{208}$ arises from the total contribution of  $DD$ and $DD^*$ loops, depicted in subfigures (a) for $\Lambda_D=2$~GeV, and (b) for $\Lambda_D=3$~GeV.}
	\label{FIG_r_total}
\end{figure*} 

\begin{table}[h]
    \centering
    \caption{The $J/\psi$ mass shift in nuclei $\text{O}^{16}$, $\text{Ca}^{40}$, $\text{Zr}^{90}$, and $\text{Pb}^{208}$ arises from the total contribution of  $DD$ and $DD^*$ loops for various values of $r$ and $\Lambda_D$.}
 
    \footnotesize
    \begin{tabular}{|c|c|c|c|}
        \hline
        \multirow{2}{*}{\textbf{A}} & \multirow{2}{*}{\textbf{r (fm)}} & \multicolumn{2}{c|}{\textbf{\shortstack{$\Delta m_{\psi}$ (MeV) \\ \scriptsize for $DD + DD^*$ loops}}} \\ 
        \cline{3-4}
         & & $\Lambda_D=2$~GeV & $\Lambda_D=3$~GeV \\ 
        \hline

        \multirow{4}{*}{$^{16}_{8}$O}  
        & 0  & -7.53  & -14.2 \\
        \cline{2-4}
        & 2  & -6.74      &  -12.7    \\
        \cline{2-4}
        & 4  &  -0.63     &   -1.22   \\
        
        \hline

        \multirow{4}{*}{$^{40}_{20}$Ca} 
        & 0  & -7.53      & -14.2      \\
        \cline{2-4}
        & 2  &  -7.26     &  -13.7    \\
        \cline{2-4}
        & 4  &   -2.42    &  -4.7    \\
        
        \hline

        \multirow{4}{*}{$^{90}_{40}$Zr} 
        & 0  &  -7.50     &    14.1  \\
        \cline{2-4}
        & 2  &   -7.48    &  -14.1    \\
        \cline{2-4}
        & 4  &   -6.46    & -12.2     \\
        \cline{2-4}
        & 6  &   -0.70    &     -1.37 \\
        \hline

        \multirow{4}{*}{$^{208}_{82}$Pb} 
        & 0  &  -7.45     & -14.0    \\
        \cline{2-4}
        & 2  &   -7.45    &   -14.0   \\
        \cline{2-4}
        & 4  &   -7.43    &  -14    \\
        \cline{2-4}
        & 6  &  -6.54     & -12.4     \\
        \hline
    \end{tabular}
    \label{table_mass_r}
\end{table}
% Now, we examine the behavior of the $J/\psi$ meson when it is captured by a nucleus.
% As discussed previously, the mass shift of the $J/\psi$ meson is directly dependent on both the real and complex components of the optical potential, as demonstrated in Eqs. \ref{eq_real} and \ref{eq_img}. This dependency is crucial for calculating the binding energy and absorption decay width ($\Gamma$) of the $J/\psi$ meson.
Now, we present the results of the $J/\psi$ mass shift as a function of radial distance $r$ from the center of the nuclei $\text{O}^{16}$, $\text{Ca}^{40}$, $\text{Zr}^{90}$, and $\text{Pb}^{208}$, evaluated at various values of the cutoff parameter $\Lambda_D$. These results include contributions from the $DD$ and $DD^*$ loop interactions as depicted in Figs. \ref{FIG_r_diff} and \ref{FIG_r_total}. As seen in Fig. \ref{FIG_r_diff}(a), at $\Lambda_D=2$~GeV, a reduction in the magnitude of $\Delta m_{\psi}$ is observed as one moves away from the center of the nucleus, from the contribution of $DD$ loop interactions across all nuclei. It is noted that heavy nuclei exhibit a larger negative $J/\psi$ mass shift compared to lighter nuclei. This effect becomes more pronounced for $DD^*$ loop interactions, as depicted in Fig. \ref{FIG_r_diff}(b). Furthermore, an increase in the cutoff parameter $\Lambda_D$ leads to a more substantial downward shift in $J/\psi$ mass within finite nuclei, as shown in Fig. \ref{FIG_r_diff}(c) and (d). This pattern becomes more evident in Fig. \ref{FIG_r_total}, which illustrates the mass shift of the $J/\psi$ mesic-nuclei, resulting from the combined effects of $DD$ and $DD^*$ loop interactions. The results of $J/\psi$ mass shift in nuclei $\text{O}^{16}$, $\text{Ca}^{40}$, $\text{Zr}^{90}$, and $\text{Pb}^{208}$ arises from the total contribution of  $DD$ and $DD^*$ loops for various values of $r$ and $\Lambda_D$ are listed in table \ref{table_mass_r}.  
\begin{table}[]
    \centering
    \caption{The binding energy $E$ and absorption decay width $\Gamma$ of $J/\psi$ meson for both the ground and excited states of $\text{O}^{16}$, $\text{Ca}^{40}$, $\text{Zr}^{90}$, and $\text{Pb}^{208}$ nuclei (in units of MeV)}
\begin{tabular} {|c|c|c|c|c|c|c|c|}
\hline
\multirow{2}{*}{A}  ~& \multirow{2}{*}{$nl$}  ~& \multicolumn{3}{c|}{$E$ (MeV)}  ~& \multicolumn{3}{c|}{$\Gamma$ (in MeV)} \\ 
\cline{3-8}
  ~&  ~& $\kappa$ =0  ~& $\kappa$ = 0.5  ~& $\kappa$ =1  ~& $\kappa$ =0  ~& $\kappa$ =  0.5  ~& $\kappa$ =1 \\
\hline
\multirow{1}{*}{$^{16}_{8}$O}  ~& $1s$   ~& -7.827   ~& -7.745   ~& -7.521  ~& 0.0  ~& 6.121  ~& 12.371  \\
\cline{2-8}
 ~& $1p$   ~& -2.133   ~& -1.925   ~& -1.391  ~& 0.0  ~& 4.264  ~& 8.91  \\
\hline
\multirow{1}{*}{$^{40}_{20}$Ca}  ~& $1s$   ~& -8.563   ~& -8.513   ~& -8.374  ~& 0.0  ~& 5.626  ~& 11.318  \\
\cline{2-8}
 ~& $1p$   ~& -5.054   ~& -4.967   ~& -4.726  ~& 0.0  ~& 4.56  ~& 9.244  \\
\cline{2-8}
 ~&  $1d$   ~& -1.903   ~& -1.763   ~& -1.39  ~& 0.0  ~& 3.395  ~& 7.012  \\
\cline{2-8}
 ~&  $2s$   ~& -2.335   ~& -2.19   ~& -1.801  ~& 0.0  ~& 3.322  ~& 6.873  \\
\hline
\multirow{1}{*}{$^{90}_{40}$Zr}  ~& $1s$   ~& -9.982   ~& -9.946   ~& -9.846  ~& 0.0  ~& 6.015  ~& 12.076  \\
\cline{2-8}
 ~& $1p$   ~& -7.306   ~& -7.246   ~& -7.078  ~& 0.0  ~& 5.265  ~& 10.611  \\
\cline{2-8}
 ~&  $1d$   ~& -4.773   ~& -4.685   ~& -4.442  ~& 0.0  ~& 4.488  ~& 9.100  \\
\cline{2-8}
 ~&  $2s$   ~& -4.963   ~& -4.872   ~& -4.624  ~& 0.0  ~& 4.459  ~& 9.046  \\
\cline{2-8}
 ~&  $2p$   ~& -2.807   ~& -2.68   ~& -2.335  ~& 0.0  ~& 3.603  ~& 7.402  \\
\cline{2-8}
 ~&  $2d$   ~& -0.958   ~& -0.77   ~& -0.291  ~& 0.0  ~& 2.635  ~& 5.598  \\
\hline
\multirow{6}{*}{$^{208}_{82}$Pb}  ~& $1s$   ~& -11.146   ~& -11.121   ~& -11.05  ~& 0.0  ~& 6.309  ~& 12.649  \\
\cline{2-8}
 ~& $1p$   ~& -9.231   ~& -9.19   ~& -9.075  ~& 0.0  ~& 5.799  ~& 11.653  \\
\cline{2-8}
 ~&  $1d$   ~& -7.367   ~& -7.309   ~& -7.148  ~& 0.0  ~& 5.282  ~& 10.641  \\
\cline{2-8}
 ~&  $2s$   ~& -7.439   ~& -7.38   ~& -7.216  ~& 0.0  ~& 5.271  ~& 10.622  \\
\cline{2-8}
 ~&  $2p$   ~& -4.08   ~& -3.98   ~& -3.705  ~& 0.0  ~& 4.172  ~& 8.49  \\
\hline
\end{tabular}
 
    \label{table_BEW}
\end{table}
For the $J/\psi$ meson, we observed a negative mass shift, indicating the potential formation of a meson-nucleus bound state. The calculated binding energy and absorption decay width for both the ground and excited states of nuclei are detailed in Table \ref{table_BEW}. In the case of the light nuclei $\text{O}^{16}$ nucleus, the bound states with negative binding energy are restricted to the $1s$ and $1p$ orbitals for phenomenological parameters $\kappa=0$, 0.5, and 1. As $\kappa$ approaches unity, the magnitude of binding energy of meson is observed to decrease while its decay width values increase. This implies that the absorption of mesons in the nucleus is greater at higher $\kappa$ values, which may lead to peak broadening and hinder the clear identification of bound states.  The $\text{Ca}^{40}$ nucleus exhibits stability for the $1s$, $1p$, $1d$, and $2s$ bound states, whereas $\text{Zr}^{90}$, and $\text{Pb}^{208}$ support a broader range of bound states, extending from $1s$ to $2d$ and $2p$, respectively, as illustrated in Table \ref{table_BEW}. These findings indicate that the greater the magnitude of the attractive potential in heavy nuclei, the more significantly stable states are expected. The $J/\psi$-nucleus system is notably stable, as its energy lies below the nucleon knockout threshold. The exothermic reaction $J/\psi N \to \eta_c N$ results in a minimal decay width, allowing this bound state to remain sufficiently narrow for experimental observation \cite{Sibirtsev:2000aw}. In contrast to the $\eta$ and $\omega$ mesons within different nuclei \cite{Tsushima:1998qw}, which experience significant broadening due to strong interactions, the $J/\psi$ can be captured into discrete bound states through recoilless kinematics. For example, as discussed in Ref. \cite{Tsushima:1998qw}, the $\eta$ meson in $\text{O}^{16}$ ($\text{Pb}^{208}$) nuclei exhibits a binding energy of $-32.6$ ($-56.3$) MeV and a decay width of $26.7$ ($33.2$) MeV. Similarly, the $\omega$ meson binds at $-46$ ($-118$) MeV with a corresponding decay width of $31.7$ ($33.1$) MeV.
Earlier works, including the Refs. \cite{Tsushima:2011kh} and \cite{CobosMartinez:2021bia} have calculated the binding energies of $J/\psi$-nuclei using the QMC model, their corresponding values are shown in Table \ref{tab:bound-state_ref} and are compared with our findings. Moreover, medium and heavy nuclei are expected to show a rich range of bound state spectra. However, experimental studies depend on producing $J/\psi$ mesons with minimal relative momentum to the nucleus. By carefully choosing the beam and target nuclei, it may be possible to achieve such kinematics, as current findings suggest that numerous nuclei are likely to form $J/\psi$ -nuclear bound states \cite{Tsushima:2011kh}.

\begin{table}[h!]
\centering
\caption{$J/\psi-$nucleus bound state energies for different references (in the units of MeV).}
\label{tab:bound-state_ref}
\begin{tabular}{|l|l|c|c|c|}
\hline 
A & $nl$ & \multicolumn{3}{c|}{\textbf{Bound State Energies}} \\ \cline{3-5}
 & & \textbf{Ref. \cite{Tsushima:2011kh}} & \textbf{Ref.\cite{CobosMartinez:2021bia}} & \textbf{Our results} \\ 
 & & ($\Lambda_D = 2$~GeV) & ($\Lambda_D = 3$~GeV) & ($\Lambda_D = 3$~GeV) \\ \hline

$^{40}_{J/\psi}\text{Ca}$   & 1s & -17.2 & -5.44 & -8.56  \\ \cline{2-5}
                            & 1p & 12.9 & -2.32 & -5.054 \\ \cline{2-5}
                            & 1d & -8.21 & -   & -1.90 \\ \cline{2-5}
                            & 2s & -7.48 & -   & -2.33 \\ \hline 

$^{90}_{J/\psi}\text{Zr}$   & 1s & -18.69 & -6.40 & -9.98 \\ \cline{2-5}
                            & 1p & -16.07 & -4.42& -7.30 \\ \cline{2-5}
                            & 1d & -13.06 & -2.13 & -4.77 \\ \cline{2-5}
                            & 2s & -12.22 & -1.56 & -4.96 \\ \hline 
                            
$^{208}_{J/\psi}\text{Pb}$  & 1s & -19.1 & -7.08 & -11.14 \\ \cline{2-5}
                            & 1p & -17.59 & -5.86 & -9.23 \\ \cline{2-5}
                            & 1d & -15.81 & -4.38 & -7.36 \\ \cline{2-5}
                            & 2s & -15.26 & -3.81 & -7.43 \\ \hline 
\end{tabular}
\end{table} 
 
\section{Summary} 
\label{summary}
In this work, we have examined how the contribution of $D$ and $D^*$ meson loops in an effective Lagrangian framework affects properties of $J/\psi$ meson in asymmetric nuclear matter at zero and finite temperatures. The hadronic chiral SU(3) model is employed to calculate scalar condensates, which are then incorporated in the QCD sum rules method to estimate the medium dependence of $D$ (\(D^+\), \(D^0\), \(D^{*+}\), and \(D^{*0}\)) meson masses. The in-medium reduction in masses of vector $D^*$ mesons is more significant compared to pseudoscalar $D$ mesons at finite temperature and density of medium, which is attributed to the stronger attractive interaction of the $D^*$ meson with the nuclear medium. Our analysis reveals that the $J/\psi$ meson exhibits a negative mass shift correlated with an increase in baryonic density. At a temperature of 100 MeV and isospin asymmetry $I_a=0.3$, the results derived from sum rules indicate that the mass shifts of the $D$ and $D^*$ mesons are observed to be $-2.66$ and $-4.68$ MeV, respectively. This leads to an approximate reduction of $-3.39$ and $-7.32$ MeV in the $J/\psi$ mass for the $DD$ and $ D D^*$ meson loop contributions within asymmetric nuclear matter at nuclear saturation density, with cut-off parameter $\Lambda_D=2$ GeV, respectively.
In examining the impact of the $D^* D^*$ meson loop on the $J/\psi$ self-energy, we noted a much larger negative mass shift of the $J/\psi$ meson. However, the heavier loop is expected to contribute less. 
% To ensure our minimum prediction is accurate and to minimize potential model-dependent uncertainties, we focus on the contributions from the $DD$ and $ D D^*$ meson loops. 
The observed negative mass shift suggests that nuclear mean fields exert an attractive influence on the $J/\psi$ meson, such a mass shift may be sufficient to facilitate the formation of meson-nucleus bound states. Moreover, we have computed the binding energy and absorption decay width of $J/\psi$ meson for both the ground and excited states of $\text{O}^{16}$, $\text{Ca}^{40}$, $\text{Zr}^{90}$, and $\text{Pb}^{208}$ nuclei by solving the Klein-Gordon equation. Our finding indicates that the $J/\psi$ meson forms bound states with all the nuclei we examined, and heavier mesons exhibit a greater probability of forming a mesic-nuclei bound state. Mainly, the theoretical predictions of $J/\psi$-$\text{Pb}^{208}$ bound state strongly encourage experimental studies of nuclei after the upgraded 12 GeV CEBAF at JLab facility \cite{Adderley:2024czm}. This study may serve as a useful reference for experimental studies of quarkonium suppression in heavy-ion collisions and medium effects through the formation of bound states of mesons, associated with the FAIR project, such as CBM and PANDA \cite{senger2012compressed, Agarwal:2023otg, Durante:2019hzd}, and J-PARC collaboration \cite{Tanida:2016ryv}, and CEBAF experiment at JLab  \cite{Adderley:2024czm}.  
% The experimental search for the //^P-nuclear bound 
% states will be possible at JLab after 12 GeV upgrade of CEABAF. 
% % the NA38, and NA60 collaborations at CERN SPS  \cite{NA38:1989yxm, NA38:1994yau, NA60:2007odv}, the PHENIX and ALICE experiments at RHIC \cite{PHENIX:2006gsi, PHENIX:2011img, ALICE:2012jsl}.
% For $ D^*$ meson loop contribution on the $J/\psi$ self-energy, we observed a large mass shift of $J/\psi$, the heavier loop should give a lesser contribution, however. To ensure the accuracy of our
% minimum prediction and to avoid possible model-dependent uncertainties, we concentrate on the $DD$ and $ D^* D^*$  meson-loop contributions. 
% These theoretical understandings of the in-medium behavior of heavy-flavor mesons are significant to analyzing medium effects and hadronization phenomena explored in Jefferson Lab experiments \cite{Adderley:2024czm, GlueX:2023pev}. 
% These data will allow CBM to successfully explore the onset of chiral symmetry
% restoration and its effect on dilepton production, exotic forms of (strange) QCD matter
% and the equation-of-state at condition close to those encountered in neutron star
\section*{\bf Acknowledgement}
The authors sincerely acknowledge the support for this
work from the Ministry of Science and Human Resources
(MHRD), Government of India, through an Institute fellowship under the National Institute of Technology Jalandhar.
Arvind Kumar sincerely acknowledges Anusandhan National
Research Foundation (ANRF), Government of India, for
funding of the research project under the Science and
Engineering Research Board-Core Research Grant
(SERB-CRG) scheme (File No. CRG/2023/000557).

\bibliographystyle{elsarticle-num}   % or elsarticle-harv
\bibliography{Ref_loop}

%-------------------------------------

\end{document}